\documentclass[aps,prc,showpacs,twocolumn,nofootinbib,preprintnumbers]{revtex4-1}%preprint
\usepackage{longtable}
\usepackage{graphicx}% Include figure files
\usepackage{dcolumn}% Align table columns on decimal point
\usepackage{bm}% bold math
\usepackage{amsmath,amssymb}
\newcommand{\bra}[1]{\langle \, #1 \, |}
\newcommand{\ket}[1]{| \, #1 \, \rangle}
\newcommand{\kket}[1]{\, #1 \, \rangle}

\newcommand{\re}{\text{Re }}
\newcommand{\im}{\text{Im }}

\usepackage[normalem]{ulem}  % \sout{old text} for strikeout
\usepackage[dvips]{color} % For blue in-text comments and additions

\renewcommand\sout{\bgroup \color{red} \ULdepth=-.5ex \ULset}
\begin{document}

\preprint{YITP-17-23}


\title{Compositeness of hadron resonances in finite volume}


\author{Yujiro~Tsuchida}
\affiliation{Department of Physics, Kyoto University,
Kyoto 606-8502, Japan}
\author{Tetsuo~Hyodo}
\email[]{hyodo@yukawa.kyoto-u.ac.jp}
\affiliation{Yukawa Institute for Theoretical Physics, Kyoto University, Kyoto 606-8502, Japan}
%\thanks{}


\date{\today}

%%%%%%%%%%%%%%%%%%%%%%%%%%%%%%%%%%%%%%%%%%%%%%%%%%%%%%%%%%%%%%%%%%%%%%%%
\begin{abstract}
  We develop a theoretical framework to quantify the structure of unstable hadron resonances. With the help of the corresponding system in a finite volume, we define the compositeness of resonance states which can be interpreted as a probability. This framework is used to study the structure of the scalar mesons $f_{0}(980)$ and $a_{0}(980)$. In both mesons, the $\bar{K}K$ component dominates about a half of the wave function. The method is also applied to the $\Lambda(1405)$ resonance. We argue that a single energy level in finite volume represents the two eigenstates in infinite volume. The $\bar{K}N$ component of $\Lambda(1405)$, including contributions from both eigenstates, is found to be 58\%, and the rest is composed of the $\pi\Sigma$ and other channels.
\end{abstract}

\pacs{24.30.-v,03.65.Ge,14.20.-c}

% 24.30.-v Resonance reactions
% 03.65.Ge Solutions of wave equations: bound states
% 14.20.-c Baryons (including antiparticles)

%\keywords{}

\maketitle

%%%%%%%%%%%%%%%%%%%%%%%%%%%%%%%%%%%%%%%%%%%%%%%%%%%%%%%%%%%%%%%%%%%%%%%%
\section{Introduction}
%%%%%%%%%%%%%%%%%%%%%%%%%%%%%%%%%%%%%%%%%%%%%%%%%%%%%%%%%%%%%%%%%%%%%%%%

% general, hadron structure
It is remarkable that many new hadrons are being observed in recent high-energy experiments of hadron spectroscopy~\cite{Olive:2016xmw}. The unexpected nature of the newly observed states urges us to consider exotic configurations of hadrons, such as multiquarks, gluon hybrids, and hadronic molecules~\cite{Brambilla:2010cs,Hosaka:2016pey}. In particular, the appearance of the near-threshold states is an indication of the hadronic molecule structure, in which two or more hadrons form a loosely bound state through the hadron-hadron interactions. A classical example of the hadronic molecule is the $\Lambda(1405)$ resonance, which is considered to be a $\bar{K}N$ molecular state~\cite{Dalitz:1959dn,Dalitz:1960du,Kaiser:1995eg,Oset:1998it,Oller:2000fj,Jido:2003cb,Hyodo:2007jq,Hyodo:2011ur,Kamiya:2016jqc}. Scalar mesons near the $\bar{K}K$ threshold, $f_{0}(980)$ and $a_{0}(980)$, are also candidates of the meson-meson molecule~\cite{Weinstein:1982gc,Weinstein:1983gd,Oller:1997ti,Oller:1997ng,Oller:1998hw,GomezNicola:2001as,Baru:2003qq,Pelaez:2015qba}. To elucidate the nonperturbative dynamics of the low-energy QCD, it is desired to characterize the internal structure of hadrons in a quantitative manner.

% Compositeness
In this respect, intensive attention is paid to the compositeness of hadrons, which is defined as the overlap of the hadron wave function with the scattering states. The study of the compositeness traces back to the discussion on the field renormalization constant to reveal the composite nature of the deuteron~\cite{Weinberg:1965zz}. In a series of recent studies of the compositeness of hadrons~\cite{Baru:2003qq,Baru:2010ww,Hyodo:2011qc,Aceti:2012dd,Hyodo:2013nka,Hyodo:2013iga,Hyodo:2014bda,Sekihara:2014kya,Aceti:2014ala,Garcia-Recio:2015jsa,Guo:2015daa,Kamiya:2015aea,Sekihara:2015gvw,Guo:2016wpy,Kamiya:2016oao}, it becomes evident that there is a problem of interpretation of the compositeness of unstable states~\cite{Hyodo:2013nka,Guo:2015daa,Kamiya:2015aea,Sekihara:2015gvw,Kamiya:2016oao}. For a stable bound state, the compositeness can be interpreted as a probability of finding the molecular component in the bound state, thanks to the normalization of the wave function~\cite{Hyodo:2013nka}. On the other hand, the compositeness of an unstable resonance is in general complex, and the probabilistic interpretation is not always guaranteed. 

% Finite volume
Here we approach the interpretation problem of the compositeness of resonances from yet another viewpoint, by utilizing the finite-volume system. The complex nature of the compositeness originates in the wave function of resonances which diverges at large distance and cannot be normalized in the standard prescription. In other words, the resonance wavefunction does not belong to the standard Hilbert space~\cite{Kukulin,Moiseyev}. In contrast, in the system with a finite spatial volume, all the eigenfunctions are square integrable with a discrete eigenvalue. This motivates us to define the  compositeness of resonances using the discrete eigenstates in finite volume. General properties of the eigenstates in a finite-volume system are comprehensively studied in Refs.~\cite{Luscher:1985dn,Luscher:1986pf,Luscher:1990ux}. Recently, detailed analyses of the finite-volume energy levels in specific hadron scatterings are performed, mainly to compare with the lattice QCD data~\cite{Beane:2003da,Koma:2004wz,Lage:2009zv,Doring:2011ip,Doring:2011vk,MartinezTorres:2011pr,MartinezTorres:2012yi,Hall:2013qba,Albaladejo:2016jsg}. In particular, with the recent lattice QCD data in Ref.~\cite{Menadue:2011pd}, the structure of $\Lambda(1405)$ is discussed~\cite{Hall:2014uca,Molina:2015uqp,Liu:2016wxq}. 
The finite-volume effect can also be utilized to estimate the spatial size of hadron resonances~\cite{Sekihara:2012xp}. 

% Compositeness in finite volume / paper organization
In contrast to the previous studies, the aim of this paper is focused on the definition of the compositeness of the resonance states, by using the finite-volume system. For this purpose, we first derive the expression of the compositeness of the discrete eigenstates in finite volume in Sec.~\ref{sec:definition}. It is shown that the compositeness can be written in an analogous form with the discrete eigenstates in the infinite-volume system, but can always be interpreted as a probability. Next, in Sec.~\ref{sec:compositeness}, we propose a definition of the compositeness of resonances, by identifying the finite-volume eigenstate which represents the resonance in infinite volume. This prescription is exemplified in a single-channel scattering model with a resonance. In Sec.~\ref{sec:application}, we study the hadron-hadron systems with the new definition of the compositeness of resonances, in order to clarify the structure of the scalar mesons $f_{0}(980)$ and $a_{0}(980)$ and the $\Lambda(1405)$ resonance. A summary of this work is given in the last section. In Appendix~\ref{appendix:resonancewf}, we discuss the correspondence of the finite-volume eigenstates with the resonance state in infinite volume, using the wave function of the eigenstates in one-dimensional quantum mechanics. A general discussion on the number of the finite-volume eigenstates with respect to the resonance phenomena in infinite volume is presented in Appendix~\ref{appendix:FVstates}.

%%%%%%%%%%%%%%%%%%%%%%%%%%%%%%%%%%%%%%%%%%%%%%%%%%%%%%%%%%%%%%%%%%%%%%%%
\section{Compositeness in finite volume}\label{sec:definition}
%%%%%%%%%%%%%%%%%%%%%%%%%%%%%%%%%%%%%%%%%%%%%%%%%%%%%%%%%%%%%%%%%%%%%%%%

In this section, we derive the expression of the compositeness in a finite-volume system, using the effective field theory. We show that the compositeness can be defined for all the eigenstates in finite volume, and is always interpreted as a probability.

%------------------------------
\subsection{Effective field theory}

% Hamiltonian
The formulation of the compositeness of the discrete eigenstates in infinite volume has been given in the effective field theory framework in Refs.~\cite{Kamiya:2015aea,Kamiya:2016oao}. Here we derive the corresponding expressions in a finite-volume system. We introduce the following Hamiltonian for the description of the $s$-wave low-energy scattering of the $\psi\phi$ system to which the discrete level $B_{0}$ couples:
\begin{align}
    H
    &= 
    H_{\rm free}+H_{\rm int}
    =\int_{\Omega}d^{3}\bm{x}
    (\mathcal{H}_{\rm free}
    +
    \mathcal{H}_{\rm int}), 
    \label{eq:Hamiltonian} \\
    \mathcal{H}_{\rm free}
    &= \frac{1}{2M}\bm{\nabla}\psi^{\dag}(\bm{x})\cdot\bm{\nabla}\psi(\bm{x})
    +\frac{1}{2m}\bm{\nabla}\phi^{\dag}(\bm{x})\cdot\bm{\nabla}\phi(\bm{x}) \nonumber \\
    &\quad 
    +\frac{1}{2M_{0}}\bm{\nabla}B_{0}^{\dag}(\bm{x})\cdot\bm{\nabla}B_{0}(\bm{x})
    +\omega_{0}B_{0}^{\dag}(\bm{x})B_{0}(\bm{x}) , \\
    \mathcal{H}_{\rm int}
    &= \lambda_{0}\left(
    B_{0}^{\dag}(\bm{x})\psi(\bm{x})\phi(\bm{x})
    +\phi^{\dag}(\bm{x})\psi^{\dag}(\bm{x})B_{0}(\bm{x})
    \right) \nonumber \\
    &\quad +v_{0}\psi(\bm{x})\phi(\bm{x})\phi^{\dag}(\bm{x})\psi^{\dag}(\bm{x}) .
\end{align}
The free part of the Hamiltonian $H_{\rm free}$ contains the kinetic terms of the fields $\psi$, $\phi$, and $B_{0}$. The strengths of the contact three-point and four-point interactions are given by $\lambda_{0}$ and $v_{0}$, respectively. 

% finite volume def
In Eq.~\eqref{eq:Hamiltonian}, $\Omega$ specifies the spatial volume in which the system is defined. The infinite-volume system corresponds to $\Omega=\mathbb{R}^{3}$. Here we consider the finite-volume system in a cubic box of size $L$, namely
\begin{align}
    \Omega=\Omega_{\rm FV}\equiv[0,L]^{3} .
\end{align}
We impose the periodic boundary conditions on the fields,\footnote{Although the eigenenergies quantitatively depend on the choice of the boundary conditions, the theoretical framework in this section can be equally applied to different boundary conditions.} 
\begin{align}
    \psi(\bm{x})=\psi(\bm{x}+L\bm{n}),\quad 
    \text{etc.},
    \label{eq:bc}
\end{align}
with $\bm{n}\in \mathbb{Z}^{3}$. The field operators follow the commutation relations
\begin{align}
    [\psi(\bm{x}),\psi^{\dag}(\bm{x}^{\prime})\}
    =\delta^{3}(\bm{x}-\bm{x}^{\prime}),\quad 
    \text{etc.},
\end{align}
where $[A,B\}\equiv AB-(-)^{|B||A|}BA$ with $|A|$ being the Grassmann parity of the field $A$. In finite volume, the Fourier components of the fields are labeled by the discretized momentum $\bm{p}_{\bm{n}}=(2\pi/L) \bm{n}$:
\begin{align}
    \psi(\bm{x})
    &= \frac{1}{L^{3}}\sum_{\bm{n}}e^{i\bm{x}\cdot\bm{p}_{\bm{n}}}
    \tilde{\psi}(\bm{p}_{\bm{n}}), \quad \text{etc.} ,\\
    \tilde{\psi}(\bm{p}_{\bm{n}})
    &=\int_{\Omega_{\rm FV}} d^{3}\bm{x}e^{-i\bm{x}\cdot\bm{p}_{\bm{n}}}\psi(\bm{x}),\quad 
    \text{etc.}
\end{align}
The commutation relations are given by
\begin{align}
    [\tilde{\psi}(\bm{p}_{\bm{n}}),
    \tilde{\psi}^{\dag}(\bm{p}_{\bm{n}^{\prime}})\}
    &=L^{3}\delta_{\bm{n}\bm{n}^{\prime}},\quad 
    \text{etc.}
\end{align}
The vacuum of the system $\ket{0}$ is defined such that $\tilde{\psi}(\bm{p}_{\bm{n}})\ket{0}=\tilde{\phi}(\bm{p}_{\bm{n}})\ket{0}=\tilde{B}_{0}(\bm{p}_{\bm{n}})\ket{0}=0$.

%------------------------------
\subsection{Eigenstates and compositeness}
\label{subsec:Eigenstates}

% free eigenstates
To define the compositeness, we determine the eigenstates of the free Hamiltonian $H_{\rm free}$ and the full Hamiltonian $H$. In contrast to the infinite-volume system, all the eigenstates are the discrete levels having a real eigenvalue. The eigenstates of the free Hamiltonian, which are relevant to the present problem~\cite{Kamiya:2015aea,Kamiya:2016oao}, are given by
\begin{align}
    \ket{\bm{p}_{\bm{n}}}
    &=\frac{1}{L^{3/2}}\tilde{\psi}^{\dag}(\bm{p}_{\bm{n}})
    \tilde{\phi}^{\dag}(-\bm{p}_{\bm{n}})
    \ket{0} , \\
    \ket{B_{0}}
    &= \frac{1}{L^{3/2}}
    \tilde{B}_{0}(\bm{0})\ket{0} ,
\end{align}
and the eigenvalues are calculated as
\begin{align}
    H_{\rm free}\ket{\bm{p}_{\bm{n}}}
    &=E_{\bm{n}}\ket{\bm{p}_{\bm{n}}} , \\
    H_{\rm free}\ket{B_{0}}
    &=\omega_{0}\ket{B_{0}} ,
\end{align}
with $E_{\bm{n}}=\bm{p}_{\bm{n}}^{2}/(2\mu)$ and $\mu=mM/(m+M)$. We note that the eigenenergy of the discrete state $\omega_{0}$ is independent of the system size $L$, while all the eigenenergies $E_{\bm{n}}$ (except for $|\bm{n}|=0$) scale as $\sim L^{-2}$:
\begin{align}
    E_{\bm{n}}
    &=\frac{2\pi^{2}}{\mu L^{2}}|\bm{n}|^{2} .
    \label{eq:En}
\end{align}
Because $\ket{\bm{p}_{\bm{n}}}$ represents the two-body $\psi\phi$ system and corresponds to the continuum state in infinite volume, we refer to it as the scattering state. The $|\bm{n}|=0$ state corresponds to the scattering state with vanishing relative momentum. The eigenstates are normalized as
\begin{align}
    \bra{\bm{p}_{\bm{n}}}\kket{\bm{p}_{\bm{n}^{\prime}}}
    &=
    \delta_{\bm{n}\bm{n}^{\prime}} ,\\
    \bra{B_{0}}\kket{B_{0}}&=1 .
\end{align}
In contrast to the plane waves in infinite volume, the scattering states $\ket{\bm{p}_{\bm{n}}}$ are normalizable in finite volume. The completeness relation in this sector is given by
\begin{align}
    1
    &=
    \ket{B_{0}}\bra{B_{0}}
    +\frac{1}{L^{3}}\sum_{\bm{n}}\ket{\bm{p}_{\bm{n}}}
    \bra{\bm{p}_{\bm{n}}} ,
    \label{eq:completeness}
\end{align}
where the first (second) term is the projection onto the discrete state (scattering states).

% full eigenstates
Eigenstates of the full Hamiltonian $H=H_{\rm free}+H_{\rm int}$ are also discretized. We label the eigenstates by the index $m=0,1,2,\dotsb$ as\footnote{The eigenstates may have degeneracy due to the internal symmetries (such as spin, isospin, etc.) and to the cubic rotation symmetry.} 
\begin{align}
    H\ket{\Psi^{(m)}}
    &=
    E^{(m)}\ket{\Psi^{(m)}} ,
    \quad E^{(m+1)}\geq E^{(m)}  ,
    \label{eq:Schroedinger}
\end{align}
with the normalization condition
\begin{align}
    \bra{\Psi^{(m)}}\kket{\Psi^{(l)}}
    &=
    \delta_{ml} .
    \label{eq:normalization}
\end{align}
Because of the completeness relation in Eq.~\eqref{eq:completeness}, the eigenstate $\ket{\Psi^{(m)}}$ can be expanded by $\ket{B_{0}}$ and the scattering states as
\begin{align}
    \ket{\Psi^{(m)}}
    &=
    c^{(m)}\ket{B_{0}}
    +\frac{1}{L^{3}}\sum_{\bm{n}}\chi^{(m)}(\bm{p}_{\bm{n}})\ket{\bm{p}_{\bm{n}}} ,
    \label{eq:eigenstate}
\end{align}
with the overlap factors $\chi^{(m)}(\bm{p}_{\bm{n}})=\bra{\bm{p}_{\bm{n}}}\kket{\Psi^{(m)}}$ and $c^{(m)}=\bra{B_{0}}\kket{\Psi^{(m)}}$. We now define the compositeness $X^{(m)}$ (elementariness $Z^{(m)}$) as the overlap of $\ket{\Psi^{(m)}}$ with the scattering states (with the discrete state $\ket{B_{0}}$) as
\begin{align}
    X^{(m)}
    &=\frac{1}{L^{3}}\sum_{\bm{n}}|\chi^{(m)}(\bm{p}_{\bm{n}})|^{2} , 
    \label{eq:compositenessdef} \\
    Z^{(m)}
    &= |c^{(m)}|^{2} .
    \label{eq:elementarinessdef}
\end{align}
We note that $X^{(m)}$ and $Z^{(m)}$ can be defined for all the eigenstates $\ket{\Psi^{(m)}} $. With Eqs.~\eqref{eq:completeness} and \eqref{eq:normalization}, we can show that
\begin{align}
    X^{(m)}
    +
    Z^{(m)}
    &= 1 
    \label{eq:sumrule} ,
\end{align}
for each $m$. This guarantees that the values of the compositeness and elementariness are bounded as $0\leq X^{(m)}\leq 1$ and $0\leq Z^{(m)}\leq 1$, and they are interpreted as probabilities. We emphasize that the standard normalization in Eq.~\eqref{eq:normalization} is essential for the probabilistic interpretation. The unstable states cannot be normalized in this form, and the use of the biorthogonal basis leads to the complex compositeness~\cite{Hyodo:2013nka}.

%------------------------------
\subsection{Closed-form expressions}

% expressions
In the present framework, the Schr\"odinger equation can be exactly solved, so that the compositeness and elementariness in Eqs.~\eqref{eq:compositenessdef} and \eqref{eq:elementarinessdef} are written in a closed form. Using the expansion~\eqref{eq:eigenstate}, the Schr\"odinger equation~\eqref{eq:Schroedinger} can be expressed as a coupled-channel equation for $c^{(m)}$ and $\chi^{(m)}(\bm{p}_{\bm{n}})$. By eliminating $c^{(m)}$, we obtain the equation for $\chi^{(m)}(\bm{p}_{\bm{n}})$ as
\begin{align}
    (E_{\bm{n}}-E^{(m)})\chi^{(m)}(\bm{p}_{\bm{n}})
    +v(E^{(m)})\frac{1}{L^{3}}
    \sum_{\bm{n}^{\prime}}\chi^{(m)}(\bm{p}_{\bm{n}^{\prime}})
    &= 0 ,
    \label{eq:chiequation}
\end{align}
with
\begin{align}
    v(E)
    &= 
    v_{0}+\frac{\lambda_{0}^{2}}{E-\omega_{0}} .
\end{align}
Solving Eq.~\eqref{eq:chiequation} for $\chi^{(m)}(\bm{p}_{\bm{n}})$ with the help of Eq.~\eqref{eq:sumrule}, we can express the compositeness and the elementariness as
\begin{align}
    X^{(m)}
    &= 
    \frac{I_{\rm FV}^{\prime}(E^{(m)})}
    {I_{\rm FV}^{\prime}(E^{(m)})-[1/v(E^{(m)})]^{\prime}}
    \label{eq:compositeness} ,
    \\
    Z^{(m)} &= 
    \frac{-[1/v(E^{(m)})]^{\prime}}
    {I_{\rm FV}^{\prime}(E^{(m)})-[1/v(E^{(m)})]^{\prime}}
    \label{eq:elementariness} ,
\end{align}
with\footnote{Note that the infinite series in Eq.~\eqref{eq:IFV} does not converge. In the following, it is implicit that either the divergence at large $|\bm{n}|$ is properly regularized~\cite{Beane:2003da}, or Eq.~\eqref{eq:IFV} is understood as the analytic continuation of the generalized $\zeta$ function~\cite{Luscher:1990ux}. In both cases, its derivative $I_{\rm FV}^{\prime}(E)$, which is used in the definition of the compositeness, is convergent and gives the same result.}
\begin{align}
    I_{\rm FV}(E)
    &= 
    \frac{1}{L^{3}}\sum_{\bm{n}}
    \frac{1}{E-E_{\bm{n}}} ,
    \label{eq:IFV}
\end{align}
and $A^{\prime}=dA/dE$. It can be shown from Eq.~\eqref{eq:chiequation} that $E^{(m)}$ satisfies
\begin{align}
    1-I_{\rm FV}(E^{(m)})v(E^{(m)})
    &= 
    0 .
    \label{eq:IV1}
\end{align}
Thus, the eigenenergy is determined by solving Eq.~\eqref{eq:IV1}. In the non-interacting limit ($v\to 0$), this condition means
\begin{align}
    I_{\rm FV}(E^{(m)})
    &\to
    \infty ,
\end{align}
which is satisfied by the eigenenergy of the free Hamiltonian $E_{\bm{n}}$, as easily verified from the definition in Eq.~\eqref{eq:IFV}. Because the function $I_{\rm FV}(E)$ depends on the box size $L$, the values of the eigenenergy $E^{(m)}$ and the compositeness $X^{(m)}$ depend on $L$. In Sec.~\ref{subsec:prescription}, we discuss the prescription to define the compositeness of resonances by the $L$ dependence of these quantities. 

% comments
It is instructive to compare the results with those in infinite volume in Refs.~\cite{Kamiya:2015aea,Kamiya:2016oao}. The compositeness~\eqref{eq:compositeness} and the elementariness~\eqref{eq:elementariness} can formally be obtained by replacing the loop function $G(E_{h})$ by $I_{\rm FV}(E^{(m)})$ in the corresponding expressions in infinite volume, where $E_{h}$ is a discrete eigenenergy. There, the functions $v(E)$ and $G(E)$ can be regarded as the interaction kernel and the loop function in the scattering amplitude 
\begin{align}
    T(E)=[1/v(E)-G(E)]^{-1} .
    \label{eq:amplitudeNR}
\end{align}
The condition for the eigenenergy~\eqref{eq:IV1} can be obtained from the pole condition of $T(E)$ with the same replacement. An equivalent expression with Eq.~\eqref{eq:compositeness} was introduced in Ref.~\cite{Molina:2015uqp}, which was conjectured as the compositeness without the derivation. Here we explicitly derive this expression from the overlap with the wave function~\eqref{eq:compositenessdef}.

% generalization: coupled-channel, relativistic kinematics
This framework can be generalized to the system coupled with $N$ two-body channels as in Refs.~\cite{Kamiya:2015aea,Kamiya:2016oao}. By introducing channel index $i=1,\dots,N$, the compositeness in channel $i$ is defined as
\begin{align}
    X^{(m)}_{i}
    &=\frac{1}{L^{3}}\sum_{\bm{n}}|\chi_{i}^{(m)}(\bm{p}_{\bm{n}})|^{2} , 
    \label{eq:compositenessdefmulti} 
\end{align}
where $\chi_{i}^{(m)}(\bm{p}_{\bm{n}})$ is the overlap with the scattering state in channel $i$. The wave function of an eigenstate is decomposed into the compositeness $X^{(m)}_{i}$ and the elementariness $Z^{(m)}$ as 
\begin{align}
    \sum_{i}^{N}X^{(m)}_{i}
    +
    Z^{(m)}
    &= 1 
    \label{eq:sumrulemulti} .
\end{align}
Note that the contribution of the missing channel is included in the elementariness $Z^{(m)}$~\cite{Sekihara:2014kya,Kamiya:2016oao}. The closed-form expression of the compositeness is given by
\begin{align}
    X^{(m)}_{i}
    &= 
    \frac{I_{{\rm FV},i}^{\prime}(E^{(m)})}
    {I_{{\rm FV},i}^{\prime}(E^{(m)})-[1/v_{{\rm eff},i}(E^{(m)})]^{\prime}} ,
    \label{eq:compositenessmulti} 
\end{align}
where $I_{{\rm FV},i}(E^{(m)})$ is the function~\eqref{eq:IFV} with the replacement of $E_{\bm{n}}\to E_{\bm{n},i}$, and $v_{{\rm eff},i}(E^{(m)})$ is the effective interaction in channel $i$ obtained by the Feshbach projection method~\cite{Feshbach:1958nx,Feshbach:1962ut,Hyodo:2007jq,Kamiya:2016oao}. The eigenenergy $E^{(m)}$ is now determined by the condition
\begin{align}
    \det[1-I_{\rm FV}(E^{(m)})v(E^{(m)})]
    &= 0  ,
\end{align}
with the diagonal matrix $I_{\rm FV,i}(E)$ and the coupled-channel interaction $v_{ij}(E)$~\cite{Kamiya:2016oao}. In the following sections, we utilize the scattering amplitude with relativistic kinematics. It is shown in Ref.~\cite{Sekihara:2014kya} that the compositeness is expressed by the generalization of the nonrelativistic kinematics with the suitable replacement of the loop function. Namely, as long as the scattering amplitude can be written in the form of Eq.~\eqref{eq:amplitudeNR}, the expression of the compositeness is obtained by modifying the kinematics in the loop function $G(E)$. The compositeness of the finite volume is thus obtained analogously with Eq.~\eqref{eq:compositenessmulti}, with the same generalization of the function $I_{\rm FV,i}(E)$.

%%%%%%%%%%%%%%%%%%%%%%%%%%%%%%%%%%%%%%%%%%%%%%%%%%%%%%%%%%%%%%%%%%%%%%%%
\section{Compositeness of resonances}\label{sec:compositeness}
%%%%%%%%%%%%%%%%%%%%%%%%%%%%%%%%%%%%%%%%%%%%%%%%%%%%%%%%%%%%%%%%%%%%%%%%

Here we discuss the method to define the compositeness of resonances, using the compositeness of the finite-volume eigenstates given in the previous section. In Sec.~\ref{subsec:prescription}, we present the prescription to identify the eigenstate which represents the resonance and define the compositeness of the resonance. This prescription is examined by the single-channel scattering model with one resonance in Sec.~\ref{subsec:example}.

%------------------------------
\subsection{Prescription}\label{subsec:prescription}

% intro
Let us consider an isolated resonance state in a single-channel scattering. As demonstrated in Appendix~\ref{appendix:resonancewf}, the property of the resonance is reflected in a finite-volume eigenstate when the eigenenergy is close to the resonance energy. It is therefore reasonable to consider $X^{(m)}(L)$ as the compositeness of the resonance, when the eigenenergy $E^{(m)}(L)$ is near the resonance energy.\footnote{In this subsection, we denote the $L$ dependence of $X^{(m)}$ and $E^{(m)}$ explicitly.} 

% definition of the resonance energy
The resonance energy $E_{\rm res}$ in infinite volume should however be defined carefully. First of all, $E_{\rm res}$ cannot be uniquely determined, because the finite decay width of the resonance represents the uncertainty of the energy measurement. We thus consider the following energy region
\begin{align}
    E_{\rm min} 
    &\leq E_{\rm res}
    \leq 
    E_{\rm max} 
    \label{eq:Eregion} ,
\end{align}
and regard the states satisfying this as the resonance. Next, there are two ways to determine $E_{\rm min}$ and $E_{\rm max}$. On one hand, the eigenenergy of the resonance is expressed by the pole of the scattering amplitude in the complex energy plane. In this case, we determine $E_{\rm min} = M_{\rm res}-\Gamma_{\rm res}/2$ and $E_{\rm max} = M_{\rm res}+\Gamma_{\rm res}/2$, with $M_{\rm res}$ ($-\Gamma_{\rm res}/2$) being the real (imaginary) part of the pole energy. On the other hand, the resonance energy can also be read off from the behavior of the scattering amplitude on the real axis. In this method, $E_{\rm min}$ and $E_{\rm max}$ are determined by the energies at which the spectrum (i.e., the imaginary part of the scattering amplitude) becomes a half of the peak value. If the resonance is isolated from other resonances and the nonresonant amplitude is small, then the resonant Breit-Wigner amplitude dominates so that both methods give a similar set of $(E_{\rm min} ,E_{\rm max} )$. However, this is not always the case. For instance, in the case of $\Lambda(1405)$, there are two complex poles in the relevant energy region, while there is only one peak structure in the scattering amplitude on the real axis~\cite{Jido:2003cb,Hyodo:2007jq}. As shown in Appendix~\ref{appendix:FVstates}, the finite-volume eigenenergies reflect the behavior of the phase shift on the real energy axis, rather than the poles in the complex energy plane. This suggests that the latter approach is suitable in the present purpose with the finite-volume effect to define the region of the resonance energy. Namely, we determine $(E_{\rm min} ,E_{\rm max} )$ from the behavior of the imaginary part of the scattering amplitude.

% choice of the energy level and determination of range of L
For a given set of $(E_{\rm min} ,E_{\rm max} )$, there are many states whose eigenenergy satisfies Eq.~\eqref{eq:Eregion} (see Fig.~\ref{fig:spectrasingle} in Sec.~\ref{subsec:example} and Fig.~\ref{fig:spectrum1d} in Appendix~\ref{appendix:resonancewf}). We thus need to consider the choice of the energy level to determine the compositeness. Since the infinite-volume system corresponds to the limit $L\to \infty$, one may naively think that the energy level at large $L$ should be adopted. This is however not appropriate in practice, because the energy levels become denser and denser at large $L$, and the eigenstates are largely contaminated by the scattering states. On the other hand, when we decrease the box size $L$ down to the spatial extent of the wavefunction, the finite-volume effect on the wave function largely modifies the eigenenergy~\cite{Luscher:1985dn,Koma:2004wz,Sekihara:2012xp}. Keeping these in mind, let us examine each eigenstate in finite volume. The ground state $(m=0)$ corresponds to the threshold energy $E^{(m)}(L)\sim 0$, and does not usually satisfy Eq.~\eqref{eq:Eregion}. The eigenenergy of the first excited state $E^{(1)}(L)$ can satisfy Eq.~\eqref{eq:Eregion}, when $L$ is larger than the spatial extent of the wave function and smaller than the region where the eigenenergy is affected by the lowest scattering state with a finite momentum. For higher excited states ($m>2$), Eq.~\eqref{eq:Eregion} is satisfied between two noninteracting scattering states. When the width of the resonance is small, there can be a region in which $E^{(m)}(L)$ is stable against $L$. But this is not always guaranteed, even for the resonance states with the decay width of several tens of MeV (see Fig.~\ref{fig:spectrasingle} in Sec.~\ref{subsec:example}). Thus, we shall use the first excited state ($m=1$) to determine the compositeness of the resonance. In this case, we can define the region $L_{\rm min}\leq L \leq L_{\rm max}$, where $L_{\rm min}$ ($L_{\rm max}$) is determined by the finite-volume effect on the wave function (coupling to the lowest finite energy scattering state). Of course, the value of $X^{(1)}(L)$ changes within the region $L_{\rm min}\leq L \leq L_{\rm max}$. Because the states satisfying Eq.~\eqref{eq:Eregion} are considered to represent the resonance state, we average $X^{(1)}(L)$ over the region $L_{\rm min}\leq L \leq L_{\rm max}$ in order to determine the compositeness of the resonance.

% summary
In summary, we propose the following procedure to define the compositeness of the resonance.
%\begin{itemize}
%\item 

(i) Determine $E_{\rm min}$ and $E_{\rm max}$ by the energies at which the imaginary part of the scattering amplitude in infinite volume becomes a half of the peak value.

%\item 
(ii) Determine $L_{\rm min}$ by the lower boundary where the eigenenergy of the first excited state $E^{(1)}(L)$ does not satisfy
\begin{align}
    E_{\rm min} 
    \leq E^{(1)}(L)
    \leq 
    E_{\rm max}
    \label{eq:resenergydef} ,
\end{align}
due to the finite-volume effect on the wave function.

%\item 
(iii) Determine $L_{\rm max}$ by the upper boundary where the eigenenergy of the first excited state $E^{(1)}(L)$ does not satisfy Eq.~\eqref{eq:resenergydef} due to the coupling to the lowest finite energy scattering state.

%\item
(iv) Average $X^{(1)}(L)$ in the region $L_{\rm min}\leq L\leq L_{\rm max}$ to determine the resonance compositeness $X_{\rm res}$ as
\begin{align}
   X_{\rm res} 
   &= \frac{1}{L_{\rm max}-L_{\rm min}}\int_{L_{\rm min}}^{L_{\rm max}}
   X^{(1)}(L)dL .
\end{align}
%\end{itemize}

% comment 1
Some comments are in order. First, because $X^{(1)}(L)$ is always real and positive, so is $X_{\rm res}$. In addition, because $Z^{(1)}(L)+X^{(1)}(L)=1$ is satisfied at an arbitrary $L$, we obtain the relation
\begin{align}
   X_{\rm res} + Z_{\rm res} 
   &= 1 ,
\end{align}
with 
\begin{align}
   Z_{\rm res} 
   &= \frac{1}{L_{\rm max}-L_{\rm min}}\int_{L_{\rm min}}^{L_{\rm max}}
   Z^{(1)}(L)dL .
\end{align}
Namely, $X_{\rm res}$ and $Z_{\rm res}$ can be interpreted as probabilities.

% comment 2
Second, to determine the compositeness in the above prescription, we need a theoretical model to describe the scattering amplitude. In this sense, it is important to prepare a successful model for the relevant scattering process. In general, there are two approaches to study the compositeness of hadrons. One is to use the weak-binding relation and its generalizations~\cite{Weinberg:1965zz,Kamiya:2015aea,Kamiya:2016oao}, and the other is to evaluate the compositeness at the pole position~\cite{Hyodo:2011qc,Aceti:2012dd,Sekihara:2014kya}. In the former approach, the compositeness is determined model-independently by the experimental observables, but the higher order terms in the near-threshold expansion provide some uncertainty of the result, and the applicability is limited to the near-threshold states. The latter approach determines the compositeness of any resonances without suffering from the higher order terms, but the result depends on the scattering model employed. The present procedure has similarity with the latter approach because of the use of the theoretical model of the scattering, although the compositeness is not evaluated in the complex energy plane. 

% comment 3
Third, we should keep in mind that there is no clear definition of the structure of an unstable particle. As in previous attempts~\cite{Hyodo:2013nka,Guo:2015daa,Kamiya:2015aea,Sekihara:2015gvw,Kamiya:2016oao}, the above prescription is \textit{not} an approximation of some true value of the compositeness. Rather, we propose a new plausible definition of the compositeness which can be interpreted as a probability. If the width of the resonance is too large, the present framework may not work, because of the ambiguity of the definition of the energy region~\eqref{eq:Eregion}.  This is not a limitation of the framework; it simply indicates that the ``structure'' of the resonance state with a broad width is not well defined. It is natural to expect that the compositeness is well defined only for a sufficiently narrow resonance which has a localized wave function (see Appendix~\ref{appendix:resonancewf}).

%------------------------------
\subsection{Examples in infinite volume}\label{subsec:example}

% model setup
Let us examine the above prescription by calculating the compositeness of resonances in a single-channel scattering model. We consider an $s$-wave scattering of the particles with masses $m$ and $M$. Here we adopt the relativistic kinematics where the total energy is given by $W=\sqrt{m^{2}+\bm{p}^{2}}+\sqrt{M^{2}+\bm{p}^{2}}$ for a given three-momentum $\bm{p}$. 
We construct the on-shell $T$ matrix $T(W)$ in the $N/D$ method~\cite{Chew:1960iv,Bjorken:1960zz,Oller:2000fj}, which is equivalent to the solution of the Bethe-Salpeter equation under the on-shell factorization:
\begin{align}
   T(W)
   &=[1/V(W)-G(W)]^{-1} .
   \label{eq:amplitudesingle}
\end{align}
As the interaction kernel $V(W)$, we adopt the bare-pole type interaction used in Ref.~\cite{Sekihara:2012xp}:
\begin{align}
    V(W)
    &= 
    \frac{g_{0}^{2}}{W^{2}-W_{0}^{2}} ,
    \label{eq:interactionsingle}
\end{align}
which is specified by the bare mass $W_{0}$ and the bare coupling $g_{0}$.  The loop function $G(W)$ is given by
\begin{align}
   G ( W ) 
   &= i\int
   \frac{d^{4}q}{(2\pi)^{4}}
   \frac{1}{q^{2}-m^{2}+i0^{+}}
   \frac{1}{(P-q)^{2}-M^{2}+i0^{+}} ,
   \label{eq:Gfnrel}
\end{align}
with $P^{\mu}=(W,\bm{0})$. Using the dimensional regularization, we obtain the expression
\begin{align}
   G ( W ) 
   &= \frac{1}{16 \pi ^{2}} \Bigg [ a(\mu _{\rm reg})
   + \ln \frac{mM}{\mu _{\rm reg}^{2}} 
   + \frac{M^{ 2} - m^{2}}{2 W^{2}} 
   \ln \frac{M^{ 2}}{m^{2}} \nonumber \\
   & \quad + \frac{\lambda^{1/2}}{2 W^{2}}
   \Big \{ \ln (W^{2} - m^{2} + M^{2} + \lambda^{1/2}) 
   \nonumber \\
   & \quad 
   + \ln (W^{2} + m^{2} - M^{ 2} + \lambda^{1/2}) 
   \nonumber \\ 
   & \quad 
   - \ln (- W^{2} + m^{2} - M^{ 2} + \lambda^{1/2})
   \nonumber \\ 
   & \quad 
   - \ln (- W^{2} - m^{2} + M^{ 2} + \lambda^{1/2}) \Big \}
   \Bigg ] , 
   \label{eq:loopfn}
\end{align}
where $a$ is the subtraction constant at the regularization scale $\mu_{\rm reg}$ and $\lambda=W^{4}+m^{4}+M^{4}-2W^{2}m^{2}-2m^{2}M^{2}-2M^{2}W^{2}$. For a given $W_{0}>m+M$, this model generates a resonance around $W\sim W_{0}$, unless the coupling $g_{0}$ is too large. Although the contribution from the bare state is hidden in the interaction kernel~\eqref{eq:interactionsingle}, the elementariness can be induced by the energy dependence of the interaction~\cite{Sekihara:2014kya}.

\begin{figure}[tbp]
    \centering
    \includegraphics[width=8cm]{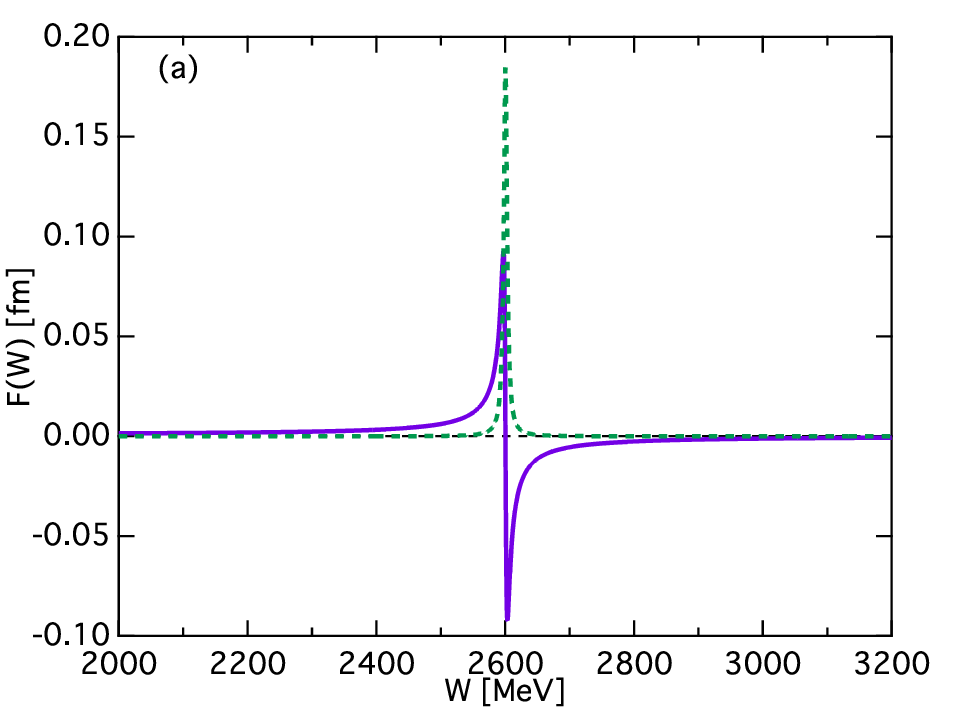}
    \includegraphics[width=8cm]{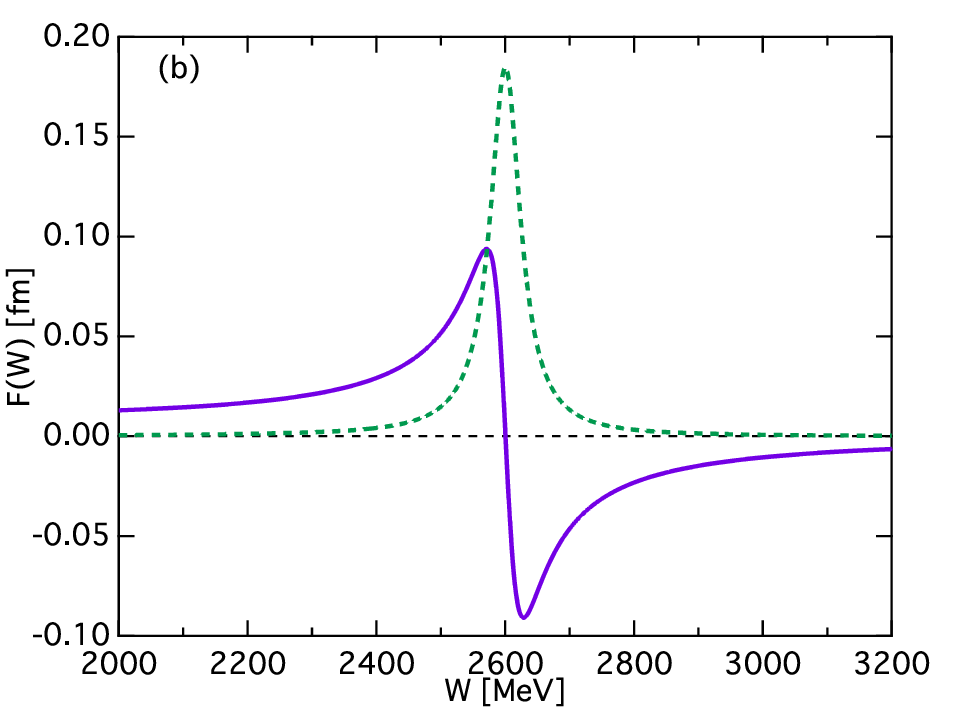}
    \includegraphics[width=8cm]{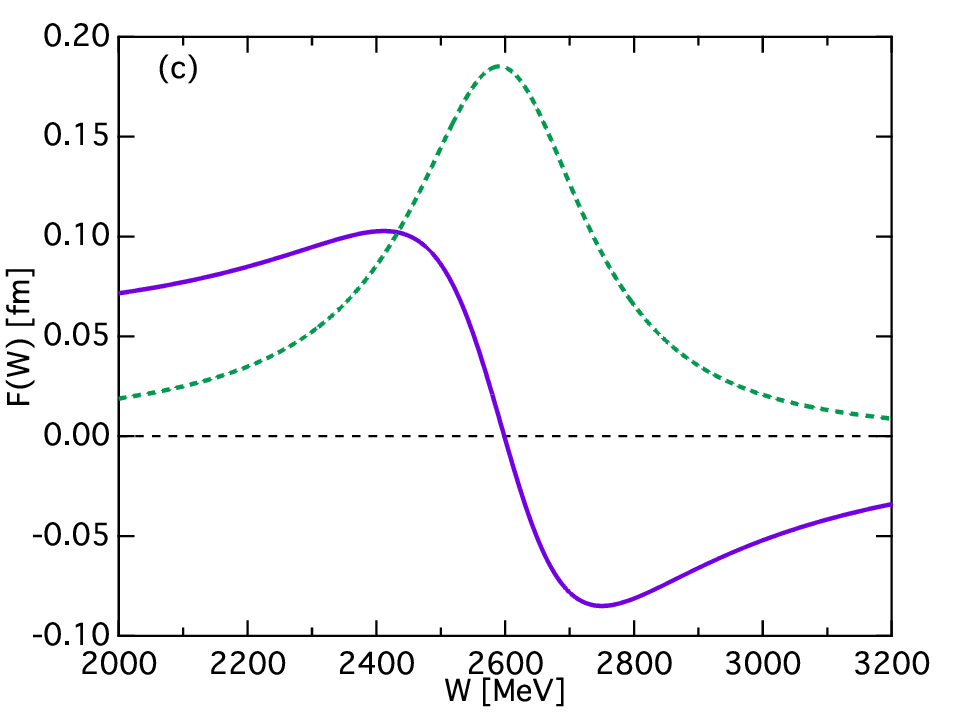}
    \caption{\label{fig:amplitudesingle}
    Real parts (solid lines) and imaginary parts (dotted lines) of the scattering amplitude $F(W)$ in model I (a), model II (b), and model III (c).}
\end{figure}

% Infinite volume results
We first calculate the scattering amplitude in infinite volume 
\begin{align}
   F(W)
   &=-\frac{1}{8\pi W}T(W) ,
\end{align}
by setting $m=495.7$ MeV, $M=938.9$ MeV, $\mu_{\rm reg}=630$ MeV, and $a(\mu_{\rm reg})=-1.95$. In order to generate a resonance around 2600 MeV, we prepare three models (I--III) with different sets of the interaction parameters $(g_{0},W_{0})$ as summarized in Table~\ref{tbl:modelsetupsingle}. The scattering amplitudes are shown in Fig.~\ref{fig:amplitudesingle}. In each model, the imaginary part of the amplitude shows a peak structure and the real part crosses zero around 2600 MeV, as a consequence of the resonance. By analytically continuing the scattering amplitude~\eqref{eq:amplitudesingle} into the complex $W$ plane, we search for the resonance pole in the second Riemann sheet. The pole positions $W_{\rm res}=M_{\rm res}-i\Gamma_{\rm res}/2$ are also summarized in Table~\ref{tbl:modelsetupsingle}. Three models correspond to a narrow width case [I, $\Gamma_{\rm res}\sim\mathcal{O}(1)$ MeV], a medium width case [II, $\Gamma_{\rm res}\sim\mathcal{O}(10)$ MeV], and a broad width case [III, $\Gamma_{\rm res}\sim\mathcal{O}(10^{2})$ several tens of MeV]. We note that a broader resonance is generated in the model with a larger bare coupling $g_{0}$, because the decay process occurs through the coupling of the bare state to the scattering state.

\begin{table}[tbp]
	\begin{center}
\caption{Interaction parameters $(g_{0},W_{0})$ and the pole positions of the infinite-volume scattering amplitude $W_{\rm res}=M_{\rm res}-i\Gamma_{\rm res}/2$ in models I--III.}
	\begin{ruledtabular}
\begin{tabular}{lccc}
Model & $g_{0}$ (MeV) & $W_{0}$ (MeV) & $W_{\rm res}$ (MeV) \\
\hline
I   & 1000 & 2600 & $2600-3i$ \\ 
II  & 3000 & 2597 & $2600-29i$ \\
III & 7000 & 2580 & $2600-165i$ \\ 
\end{tabular}
\label{tbl:modelsetupsingle}
    \end{ruledtabular}
	\end{center}
\end{table}%

%--figure---------------------------------
%--figure---------------------------------

%------------------------------
\subsection{Examples in finite volume}\label{subsec:exampleFV}

\begin{figure}[tbp]
    \centering
%    \hspace*{1cm}
    \includegraphics[width=8cm]{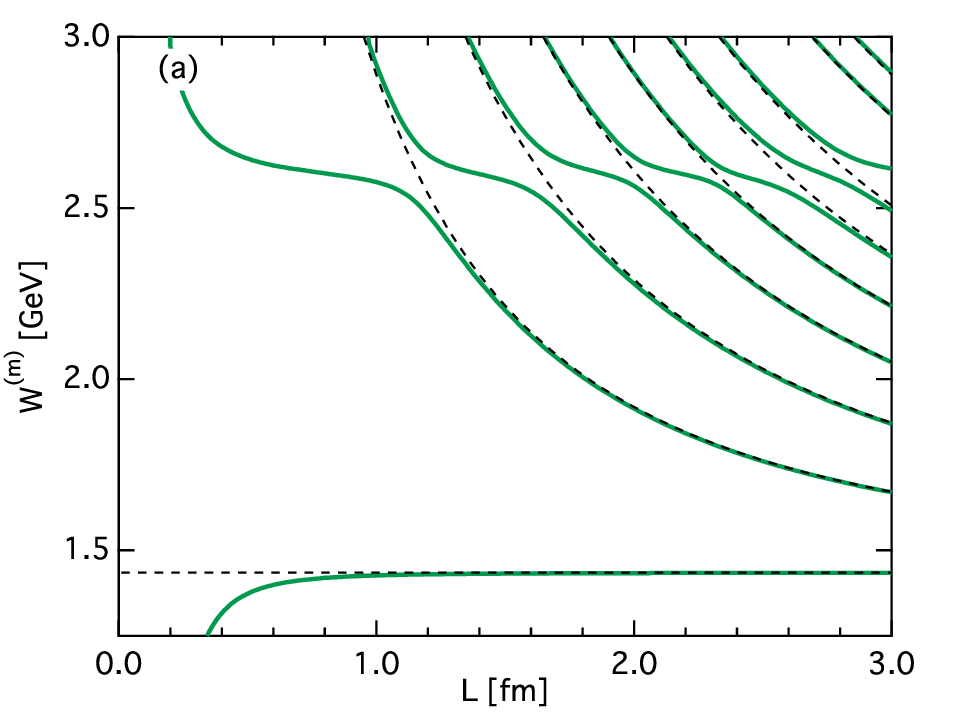}
%    \hspace*{1cm}
    \includegraphics[width=8cm]{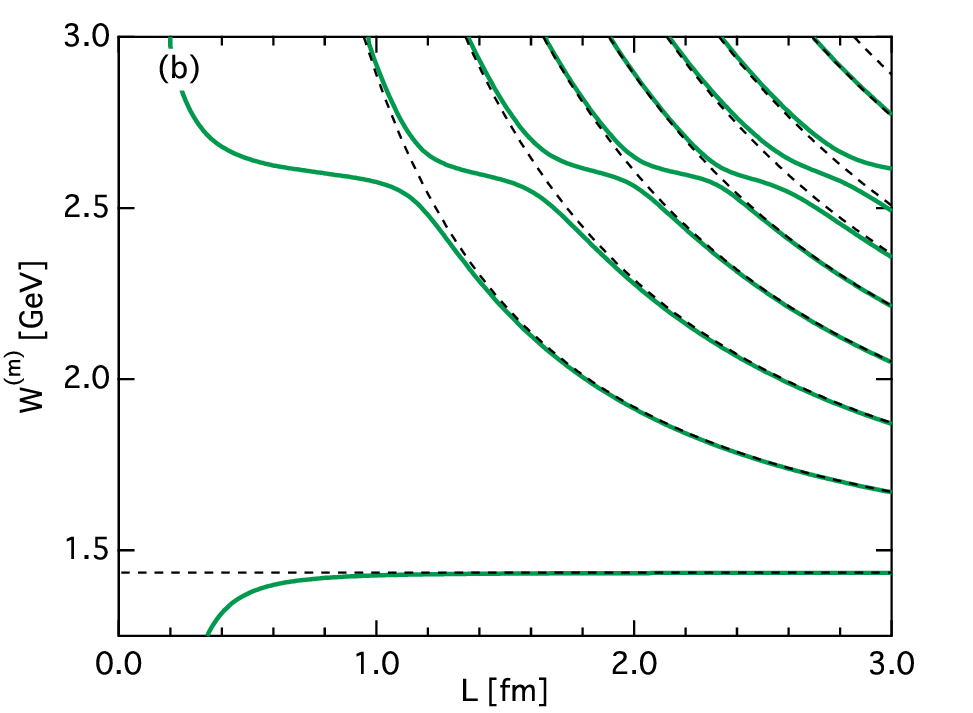}
%    \hspace*{1cm}
    \includegraphics[width=8cm]{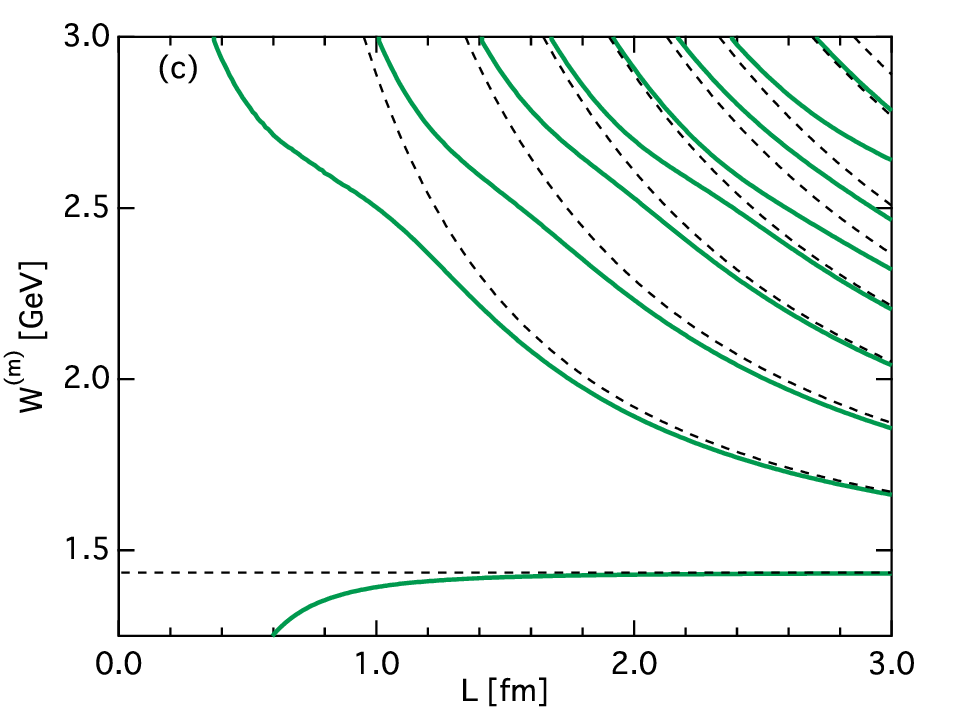}
%    \vspace*{0.5cm}
    \caption{\label{fig:spectrasingle}
    Eigenenergies $W^{(m)}(L)$ in finite volume model I (a), model II (b), and model III (c). The noninteracting eigenenergies $W_{\bm{n}}(L)$ are shown by dashed lines for comparison.}
\end{figure}

% Finite volume results
Next, we put the system in a box of size $L$ with the periodic boundary condition. The finite-volume eigenenergies $W^{(m)}$ are obtained by solving
\begin{align}
    1-G_{\rm FV}(W^{(m)})V(W^{(m)})
    &= 
    0  ,
    \label{eq:GV1}
\end{align}
with
\begin{align}
    G_{\rm FV}(W)
    &= 
    i\frac{1}{L^{3}}\sum_{\bm{n}}
    \int\frac{dq_{0}}{2\pi}
    \frac{1}{q_{0}^{2}-\bm{q}_{\bm{n}}^{2}-m^{2}+i0^{+}} 
    \nonumber \\
    &\quad \times 
    \frac{1}{(W-q_{0})^{2}-\bm{q}_{\bm{n}}^{2}-M^{2}+i0^{+}} ,
\end{align}
where $\bm{q}_{\bm{n}}=(2\pi/L) \bm{n}$. In the numerical calculation, we adopt the form introduced in Refs.~\cite{MartinezTorres:2011pr,Sekihara:2012xp}:
\begin{align}
    G_{\rm FV}(W)
    &= 
    \re [G(W)]\nonumber \\
    &\quad +\lim_{\Lambda\to\infty}
    \Biggl(
    \frac{1}{L^{3}}
    \sum_{\bm{n}}\Theta(\Lambda-|\bm{q}_{\bm{n}}|)I(W,|\bm{q}_{\bm{n}}|) \nonumber \\
    &\quad 
    -\mathcal{P}\int \frac{d^{3}\bm{q}}{(2\pi)^{3}}\Theta(\Lambda-|\bm{q}|)
    I(W,|\bm{q}|)
    \Biggr) ,\\
    I(W,q)
    &=\frac{1}{2\omega(q) E(q)}
    \frac{\omega(q)+E(q)}{W^{2}-[\omega(q)+E(q)]^{2}} , \\
    \omega (q)
    &= \sqrt{q^{2}+m^{2}},\quad
    E(q)= \sqrt{q^{2}+M^{2}} ,
\end{align}
where $\mathcal{P}$ stands for the principal value integration. The limit $\Lambda\to\infty$ is understood as the sufficiently large $\Lambda$ such that the result of $G_{\rm FV}(W)$ does not change with respect to $\Lambda$. The energy spectra of models I--III in finite volume are shown in Fig.~\ref{fig:spectrasingle} as functions of the system size $L$. The noninteracting eigenenergies $W_{\bm{n}}=\sqrt{m^{2}+\bm{p}_{\bm{n}}^{2}}+\sqrt{M^{2}+\bm{p}_{\bm{n}}^{2}}$ are also shown by dashed lines for comparison. In model I, the eigenenergies are stable against the change of the box size $L$ around the resonance energy 2600 MeV. As we have discussed in Sec.~\ref{subsec:Eigenstates}, the eigenenergy of the discrete state (scattering states) is independent of $L$ (scale as $L^{-2}$). Thus, the flat $L$ dependence of the full eigenenergy $W^{(m)}(L)$ is the indication of a narrow resonance state. In models II and III, although the eigenenergies have sizable $L$ dependence in between the scattering states, there are avoided level crossings around 2600 MeV as a consequence of the mixing of the bare state and the scattering states. In the small $L$ region, the eigenenergies deviate from 2600 MeV, due to the finite-volume effect on the wave function. In the present model, Eq.~\eqref{eq:interactionsingle} stands for the zero range interaction, and the interaction range is provided by the regularization of the loop function. As discussed in Ref.~\cite{Oller:2000fj}, $a(\mu_{\rm reg}=630 \text{ MeV})\sim -2$ corresponds to the three-momentum cutoff $\Lambda\sim 630$ MeV. We thus estimate the range of the interaction to be $1/\Lambda\sim 0.3$ fm. This roughly corresponds to the value of $L$ at which the finite-volume effect becomes prominent.

%--figure---------------------------------
%--figure---------------------------------

\begin{figure}[tbp]
    \centering
    \includegraphics[width=8cm]{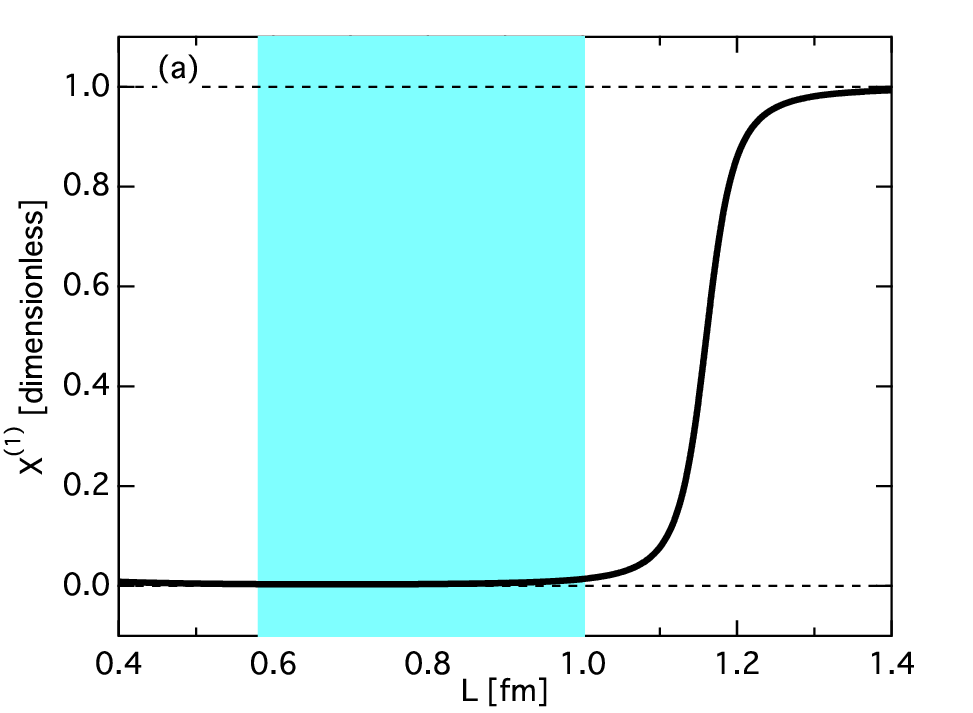}
    \includegraphics[width=8cm]{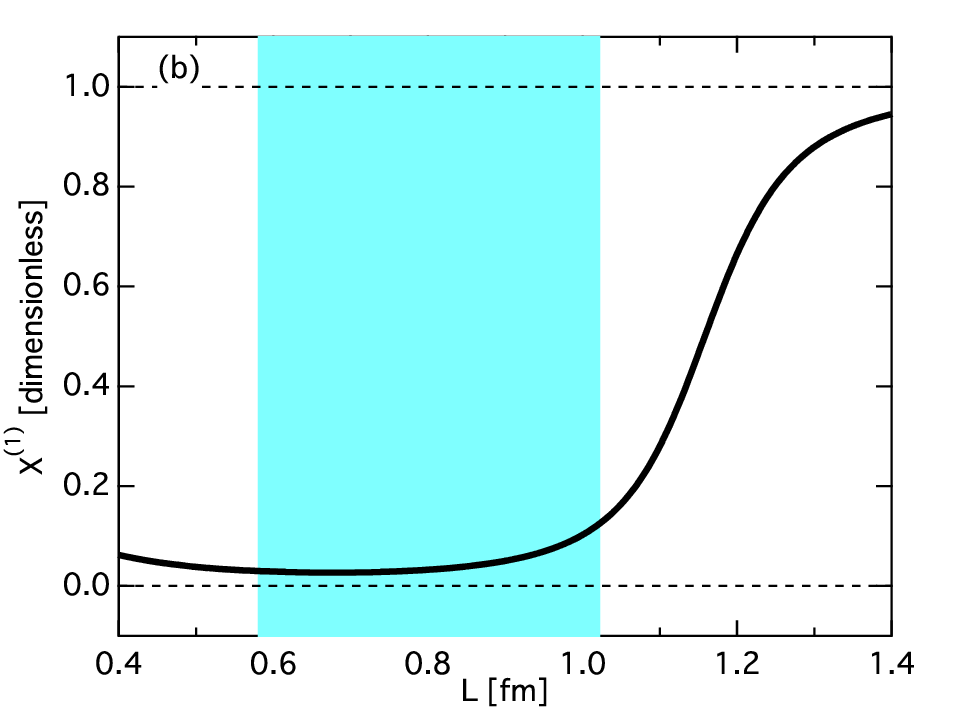}
    \includegraphics[width=8cm]{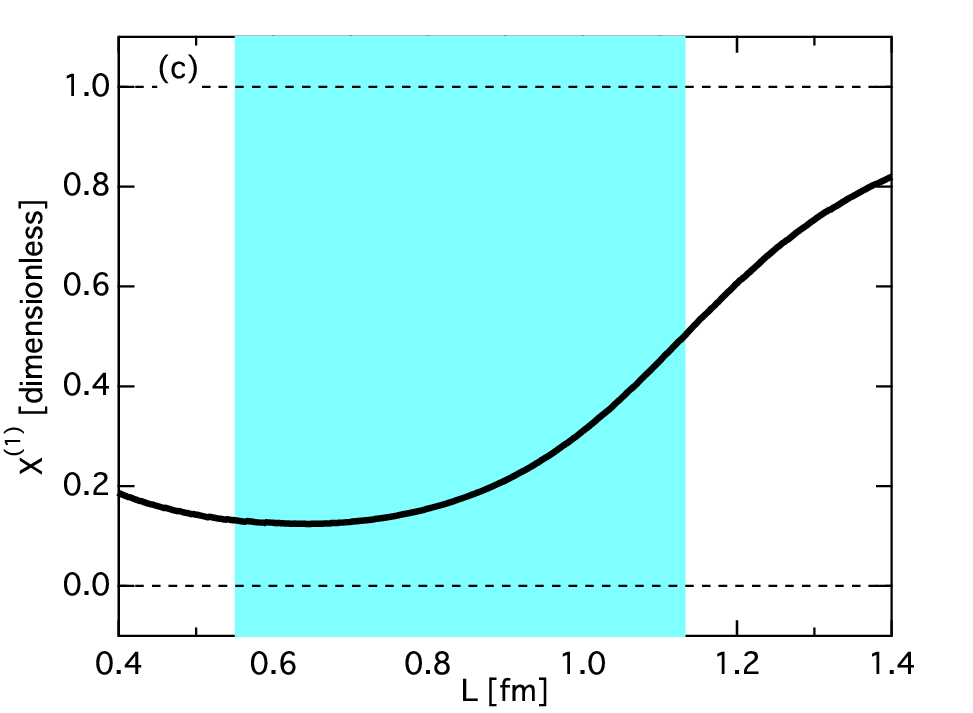}
    \caption{\label{fig:compositenesssingle}
    Compositeness of the first excited state $X^{(1)}(L)$ in 
    model I (a), model II (b), and model III (c). 
    The shaded areas represent the region $L_{\rm min}\leq L \leq L_{\rm max}$.}
\end{figure}

% compositeness
Finally, we calculate the compositeness, following the prescription presented in Sec.~\ref{subsec:prescription}. From the imaginary part of the scattering amplitude, we determine ($W_{\rm min}$, $W_{\rm max}$) as (2597.2 MeV, 2603.5 MeV) for model I, (2571.3 MeV, 2628.6 MeV) for model II, and (2415.0 MeV, 2748.6 MeV) for model III. Because the resonance is well isolated from other poles in the present models, the range of $W$ are in fair agreement with the determination by the pole position, $(M_{\rm res}- \Gamma_{\rm res}/2, M_{\rm res}+\Gamma_{\rm res}/2)$. The $L$ dependence of the energy of the first excited state then determines ($L_{\rm min}$,$L_{\rm max}$) as (0.58 fm, 1.01 fm) for model I, (0.58 fm, 1.02 fm) for model II, and (0.56 fm, 1.14 fm) for model III. In Fig.~\ref{fig:compositenesssingle}, we show the compositeness of the first excited state 
\begin{align}
    X^{(1)}(L)
    &= 
    \frac{G_{\rm FV}^{\prime}(W^{(1)})}
    {G_{\rm FV}^{\prime}(W^{(1)})-[1/V(W^{(1)})]^{\prime}} ,
\end{align}
together with the region $L_{\rm min}<L<L_{\rm max}$. In this region, the compositeness $X^{(1)}(L)$ is relatively small, indicating the elementary nature of the resonance. The increase of $X^{(1)}(L)$ at $L\sim 1.2$ fm is understood as the nature transition of the first excited state from the resonance to the scattering state, as seen in the energy spectra in Fig.~\ref{fig:spectrasingle}. Averaging $X^{(1)}(L)$ over $L_{\rm min}<L<L_{\rm max}$, we obtain the compositeness $X_{\rm res}$ as listed in Table~\ref{tbl:compositenesssingle}. In all cases, the value of the compositeness is small, indicating the importance of the bare state contribution. It can be seen that the narrower resonance has a smaller compositeness. This is intuitively understood that the state with a large fraction of the scattering channel is easy to decay.

% comparison with other works
Let us compare the results with other prescriptions of the compositeness of resonances. As shown in Refs.~\cite{Sekihara:2014kya,Guo:2015daa}, the complex-valued compositeness in infinite volume is given by 
\begin{align}
    X
    &= 
    \frac{G^{\prime}(W_{\rm res})}
    {G^{\prime}(W_{\rm res})-[1/V(W_{\rm res})]^{\prime}} 
    \label{eq:compositenessrel} .
\end{align}
The results are $X=-0.001\ 33+0.000\ 56i$ (model I), $X=-0.0123+0.0049i$ (model II), and $X=-0.0766+0.0193i$ (model III). In Refs.~\cite{Kamiya:2015aea,Kamiya:2016oao}, probabilistic interpretations of these results are presented. By defining
\begin{align}
    \tilde{X}
    &= 
    \frac{1-|1-X|+|X|}{2},
    \quad U=|X|+|1-X|-1 ,
\end{align}
it is shown that $\tilde{X}$ can be interpreted as the probability, with the uncertainty of the interpretation given by $U/2$. In Ref.~\cite{Guo:2015daa}, $X^{R}=|X|$ is shown to be interpreted as a probability, provided that the Laurent series around the resonance pole converges on the real energy axis. We show the results of $\tilde{X}\pm U/2$ and $X_{R}$ in Table~\ref{tbl:compositenesssingle}. We see that the general tendency of the results are consistent with each other, although there are some quantitative deviations. The deviation increases when the width of the resonance is large. We emphasize again that the ``true value'' of the compositeness of resonance does not exist, but the convergent result with different methods can be regarded as a measure of the structure of the resonance. In this sense, we conclude that the probabilistic interpretation is robust for a narrow width state, while the conclusion becomes ambiguous when the resonance has a broad width.

\begin{table}[tbp]
	\begin{center}
\caption{Compositeness of resonance $X_{\rm res}$ in models I-III. For comparison, we show $\tilde{X}\pm U/2$ suggested in Ref.~\cite{Kamiya:2016oao} and $X^{R}=|X|$ suggested in Ref.~\cite{Guo:2015daa}, calculated from the complex-valued compositeness at the pole position.}
	\begin{ruledtabular}
\begin{tabular}{lccc}
model & $X_{\rm res}$ & $\tilde{X}\pm U/2$~\cite{Kamiya:2016oao} & $X^{R}$~\cite{Guo:2015daa}  \\
\hline
I   & 0.0051 & $0.0001\pm 0.0014$ & $0.0014$ \\ 
II  & 0.0435 & $0.0005\pm 0.0128$ & $0.0132$ \\
III & 0.2266 & $0.0011\pm 0.0780$ & $0.0790$ \\ 
\end{tabular}
\label{tbl:compositenesssingle}
    \end{ruledtabular}
	\end{center}
\end{table}%

%--figure---------------------------------
%--figure---------------------------------

%%%%%%%%%%%%%%%%%%%%%%%%%%%%%%%%%%%%%%%%%%%%%%%%%%%%%%%%%%%%%%%%%%%%%%%%
\section{Application}\label{sec:application}
%%%%%%%%%%%%%%%%%%%%%%%%%%%%%%%%%%%%%%%%%%%%%%%%%%%%%%%%%%%%%%%%%%%%%%%%

Now we study the structure of physical hadron resonances. In Sec.~\ref{subsec:scalar}, we introduce the theoretical model to describe scalar mesons in coupled-channel meson-meson scattering based on Ref.~\cite{Sekihara:2012xp}. We then study the energy spectra in finite volume and calculate the compositeness of $f_{0}(980)$ and $a_{0}(980)$ in Sec.~\ref{subsec:scalarFV}. We perform the same analysis for $\Lambda(1405)$ in the meson-baryon scattering~\cite{Ikeda:2011pi,Ikeda:2012au}, in the infinite volume (Sec.~\ref{subsec:L1405}) and in the finite volume (Sec.~\ref{subsec:L1405FV}).

%------------------------------
\subsection{Scalar mesons in infinite volume}\label{subsec:scalar}

We consider scalar mesons in the $s$-wave meson-meson scattering amplitude around the $\bar{K}K$ threshold. In the isospin $I=0$ ($I=1$) sector, there exists $f_{0}(980)$ [$a_{0}(980)$] resonance in the $\pi\pi$-$\bar{K}K$ ($\pi\eta$-$\bar{K}K$) scattering. The meson-meson scattering has been successfully described by combining chiral perturbation theory with the unitarization scheme~\cite{Oller:1997ti,Oller:1997ng,Oller:1998hw,GomezNicola:2001as}. While there are sophisticated next-to-leading order calculations that successfully describe the phase shifts of the meson-meson scattering~\cite{Oller:1998hw,GomezNicola:2001as}, here we use a simple model with the leading order chiral Lagrangian for the interaction kernel~\cite{Sekihara:2012xp}, which reasonably well describes the experimental data and is suited to apply the present formulation of the finite volume method. In this framework, the coupled-channel scattering amplitude $T_{ij}(W)$ is obtained by Eq.~\eqref{eq:amplitudesingle} in matrix form:
\begin{align}
   T_{ij}(W)
   &=\left(\left[V^{-1}(W)-G(W)\right]^{-1}\right)_{ij} .
   \label{eq:amplitude}
\end{align}
where the indices $i,j$ represent the meson-meson channel. The interaction kernel is given by
\begin{align}
    V_{11} 
    &=\frac{m_{\pi}^2-2W^{2}}{2f^{2}} , \quad
    V_{12}=-\frac{\sqrt{3}W^{2}}{4f^{2}}, \quad
    V_{22} = -\frac{3W^{2}}{4f^{2}}
    \nonumber ,
\end{align}
for the $I=0$ channel where $i=1$ (2) corresponds to $\pi\pi$ ($\bar{K}K$) and 
\begin{align}
    V_{11} 
    &=-\frac{m_{\pi}}{3f^{2}}, \quad
    V_{12}=\frac{\sqrt{3/2}}{18f^{2}}(9W^{2}-m_{\pi}^{2}-3m_{\eta}^{2}-8m_{K}^{2}), \nonumber \\
    V_{22} 
    &= -\frac{W^{2}}{4f^{2}}
    \nonumber ,
\end{align}
for the $I=1$ channel where $i=1$ (2) represents $\pi\eta$ ($\bar{K}K$). 
The loop function matrix is given by the diagonal form
\begin{align}
    G_{ij}(W)
    &= 
    \begin{pmatrix}
    G_{1}(W) & 0 \\
    0 & G_{2}(W)
    \end{pmatrix} ,
\end{align}
where $G_{i}(W)$ is obtained as the expression in Eq.~\eqref{eq:loopfn} by adding the channel index $i$. 
The parameters are taken to be $m_{\pi}=138.0 \text{ MeV}, m_{K}=495.6\text{ MeV}, m_{\eta}=547.9\text{ MeV}$, $f=93.0$ MeV, $\mu_{\rm reg}= 1325$ MeV, and $a_{i}(\mu_{\rm reg})=-1$~\cite{Sekihara:2012xp}.

\begin{figure}[tbp]
    \centering
    \includegraphics[width=8cm]{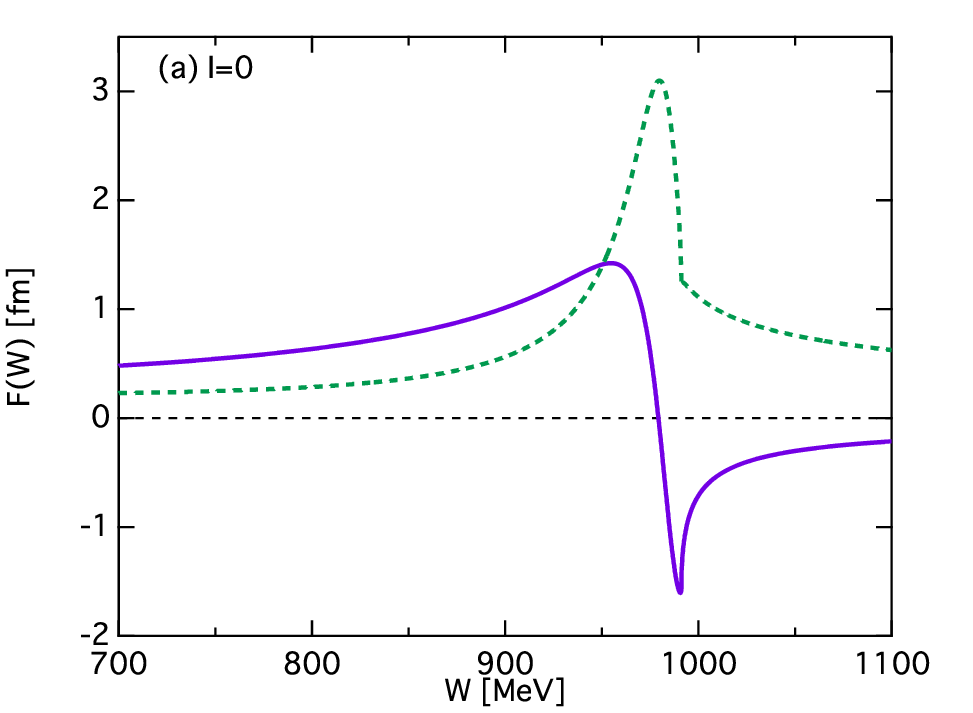}
    \includegraphics[width=8cm]{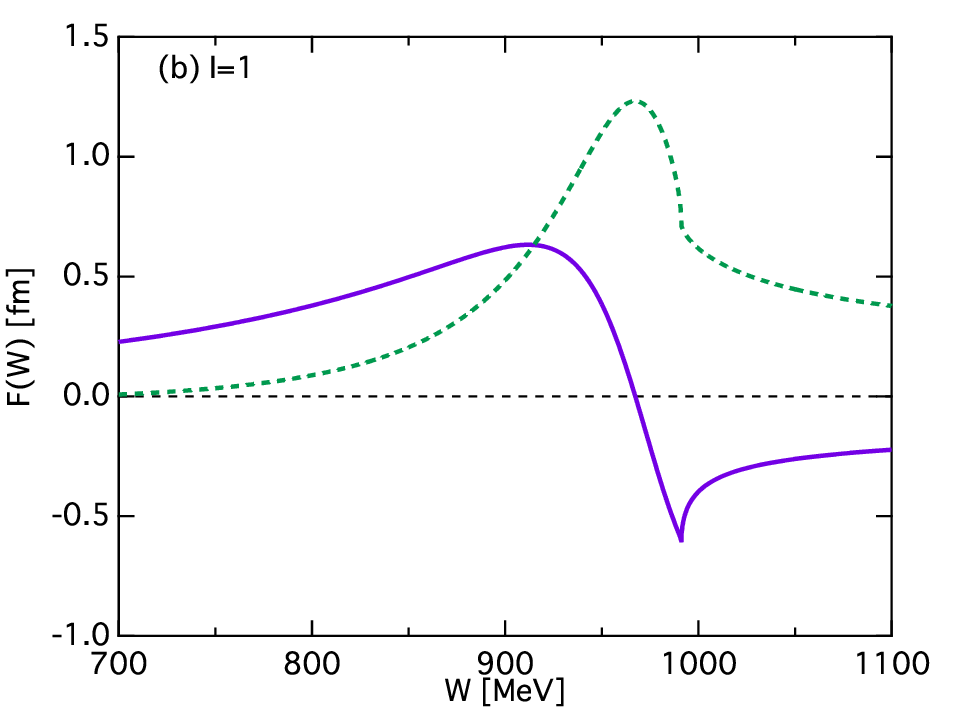}
    \caption{\label{fig:amplitude_scalar}
    Real parts (solid lines) and imaginary parts (dashed lines) of the elastic scattering amplitudes $F_{\bar{K}K}(W)$ in the $I=0$ channel (a) and in the $I=1$ channel (b).}
\end{figure}

% scattering amplitude
The scattering amplitude in the $\bar{K}K$ channel $F_{\bar{K}K}(W) =-T_{11}(W)/(8\pi W)$ is plotted in Fig.~\ref{fig:amplitude_scalar} for both $I=0$ and $I=1$ channels. Slightly below the $\bar{K}K$ threshold ($W=991.2$ MeV), a clear resonance shape is seen in each isospin channel. These correspond to the $f_{0}(980)$ and $a_{0}(980)$ resonances with $I=0$ and $I=1$, respectively. In the complex energy plane, we find a pole in the 
$I=0$ amplitude
\begin{align}
    W_{\rm res}^{I=0}
    &= 987.0-17.7i\text{ MeV} ,
\end{align}
which represents $f_{0}(980)$, and a pole in the $I=1$ amplitude
\begin{align}
    W_{\rm res}^{I=1}
    &= 979.2-53.4i\text{ MeV} ,
\end{align}
which represents $a_{0}(980)$. We note that, in the $I=0$ amplitude, there also exists a pole at $W_{\rm res}^{I=0}= 471.3-181.0i\text{ MeV}$, which represents the $f_{0}(500)$ (or $\sigma$) meson~\cite{Pelaez:2015qba}. However, the scattering amplitude does not exhibit the resonance behavior ($\pi/2$ crossing of the phase shift) near the $f_{0}(500)$ pole, and hence our method is not applicable to this state.

%--figure---------------------------------
%--figure---------------------------------

%------------------------------
\subsection{Scalar mesons in finite volume}\label{subsec:scalarFV}

\begin{figure}[tbp]
    \centering
    \includegraphics[width=8cm]{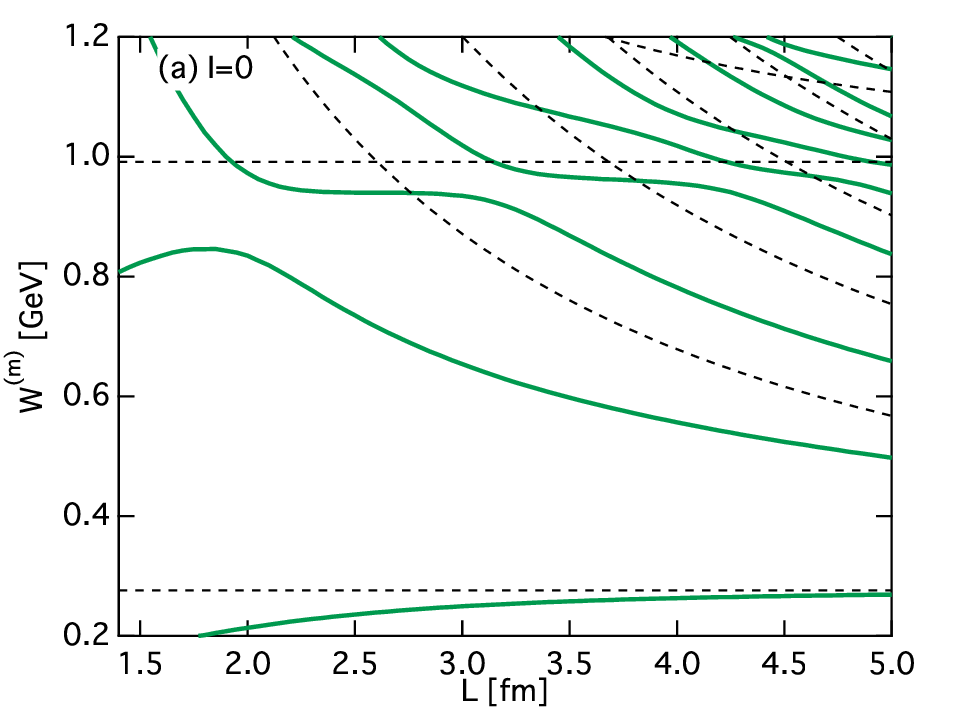}
    \includegraphics[width=8cm]{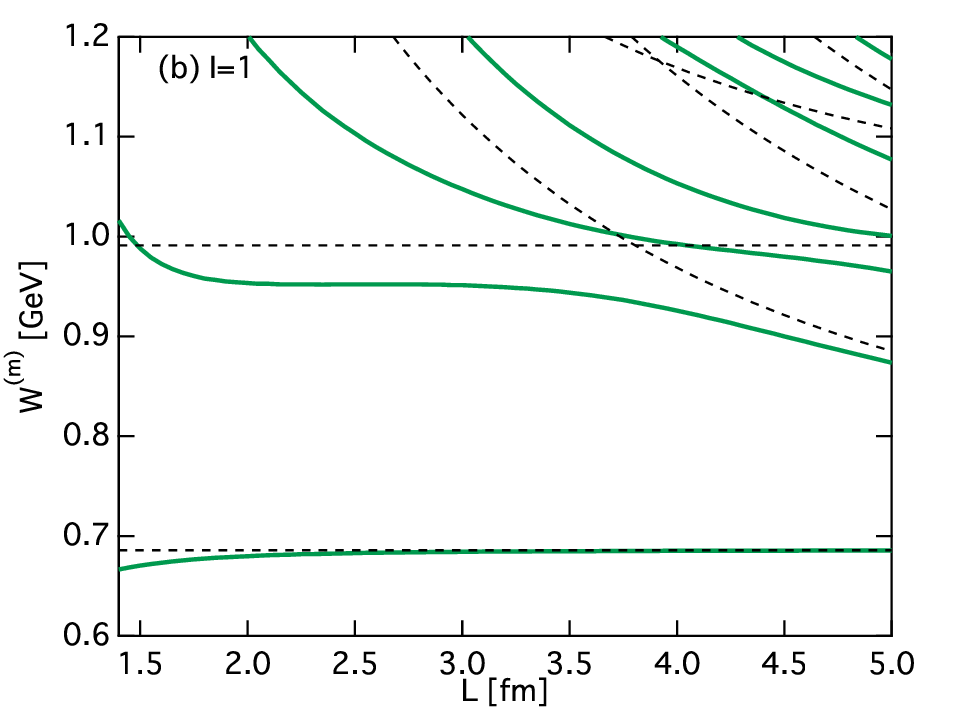}
    \caption{\label{fig:spectrum_scalar}
    Eigenenergies $W^{(m)}(L)$ in finite volume in the $\pi\pi$-$\bar{K}K$ sector in $I=0$ channel (a) and the $\pi\eta$-$\bar{K}K$ sector $I=1$ channel (b) (solid lines) in 1.4 fm $<L<5$ fm. The noninteracting eigenenergies $W_{\bm{n},1}(L)$ and $W_{\bm{n},2}(L)$ are shown by dashed lines for comparison.}
\end{figure}

We consider this model in a box of size $L$ with the periodic boundary condition. The finite-volume eigenenergies $W^{(m)}$ are determined by~\cite{MartinezTorres:2012yi,Molina:2015uqp}
\begin{align}
    \det[1-G_{\rm FV}(W^{(m)})V(W^{(m)})]
    &= 
    0 ,
\end{align}
with the diagonal matrix $\text{diag}[G_{{\rm FV},1}(W),G_{{\rm FV},2}(W)]$. The box size dependence of the eigenenergies $W^{(m)}(L)$ is shown by the solid lines in Fig.~\ref{fig:spectrum_scalar}. Shown by the dashed lines are the noninteracting eigenenergies with $V_{ij}\to 0$. Note that there are two kinds of the noninteracting energy levels, $W_{\bm{n},1}=\sqrt{m_{1}^{2}+\bm{p}_{\bm{n}}^{2}}+\sqrt{M_{1}^{2}+\bm{p}_{\bm{n}}^{2}}$ and $W_{\bm{n},2}=\sqrt{m_{2}^{2}+\bm{p}_{\bm{n}}^{2}}+\sqrt{M_{2}^{2}+\bm{p}_{\bm{n}}^{2}}$. In the $I=0$ channel, these corresponds to the $\pi\pi$ scattering states and the $\bar{K}K$ scattering states, which accumulate to the $\pi\pi$ and $\bar{K}K$ threshold energies $W_{\pi\pi}=276.0$ MeV and $W_{\bar{K}K}=991.2$ MeV in the limit $L\to \infty$, respectively. The results of the energy spectra are compatible with previous works, Figs.~1 and 2 in Ref.~\cite{Doring:2011ip} and Fig.~1 in Ref.~\cite{Doring:2011vk}. In both isospin sectors, we observe a plateau of the eigenenergy around 980 MeV, representing the scalar meson resonance. In these cases, there is one-to-one correspondence between the resonance pole in the infinite volume and the plateau of the eigenenergy in the finite volume. 

%--figure---------------------------------
%--figure---------------------------------

\begin{figure}[tbp]
    \centering
    \includegraphics[width=8cm]{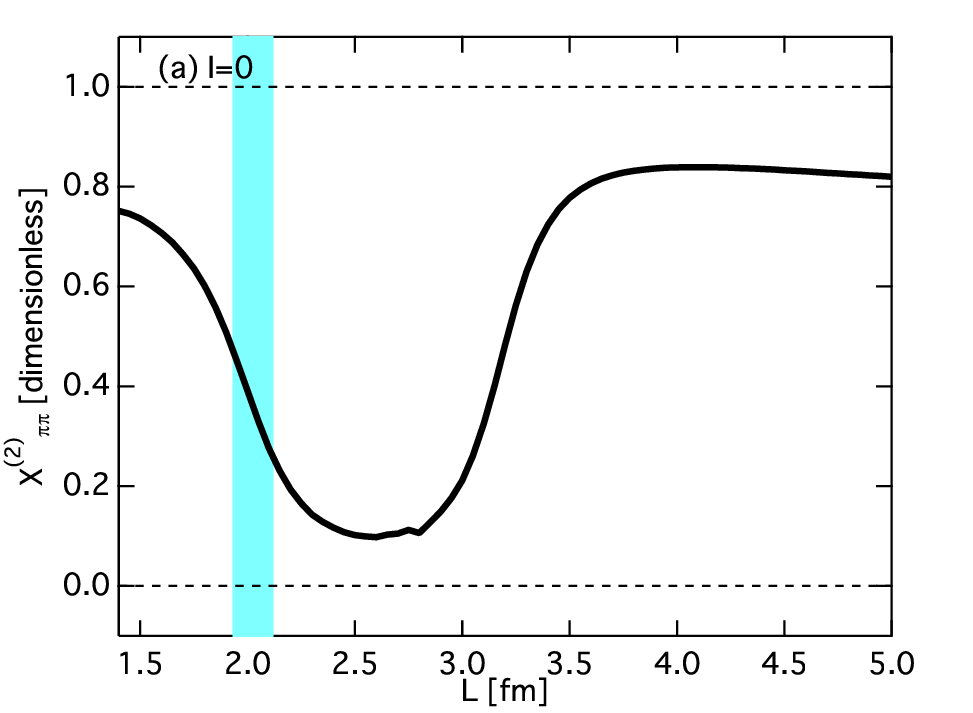}
    \includegraphics[width=8cm]{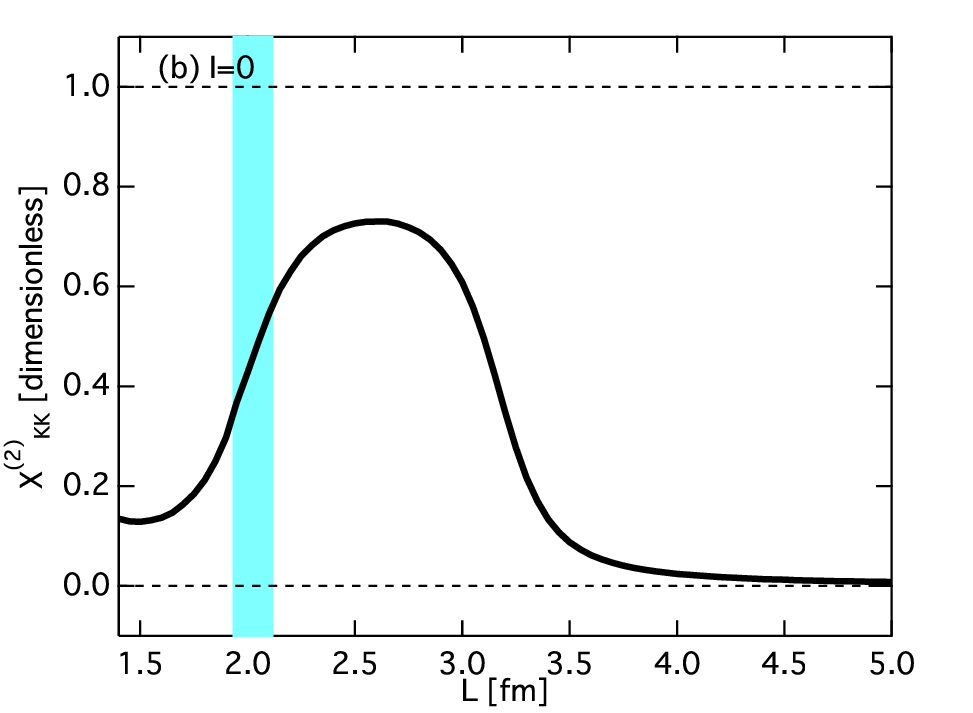}
    \includegraphics[width=8cm]{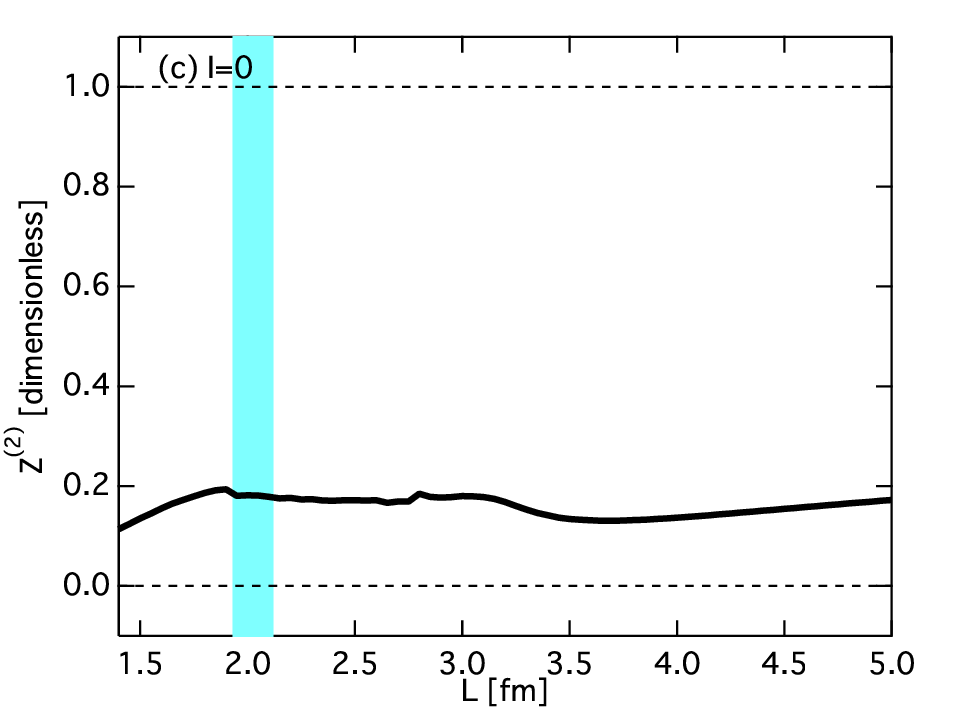}
    \caption{\label{fig:compositenessI0}
    Compositeness of the second excited state $X^{(2)}_{\pi\pi}(L)$ (a) and $X^{(2)}_{K\bar{K}}(L)$ (b) and the elementariness $Z^{(2)}(L)$ (c) in the $I=0$ scalar meson sector. The shaded areas represent the region $L_{\rm min}\leq L \leq L_{\rm max}$.}
\end{figure}

\begin{figure}[tbp]
    \centering
    \includegraphics[width=8cm]{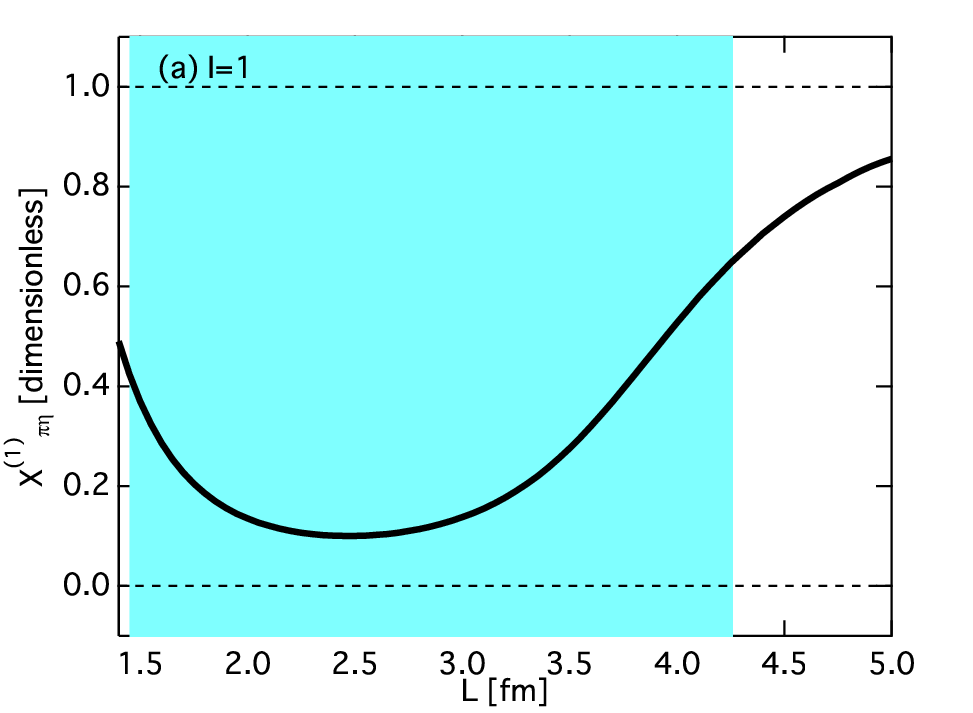}
    \includegraphics[width=8cm]{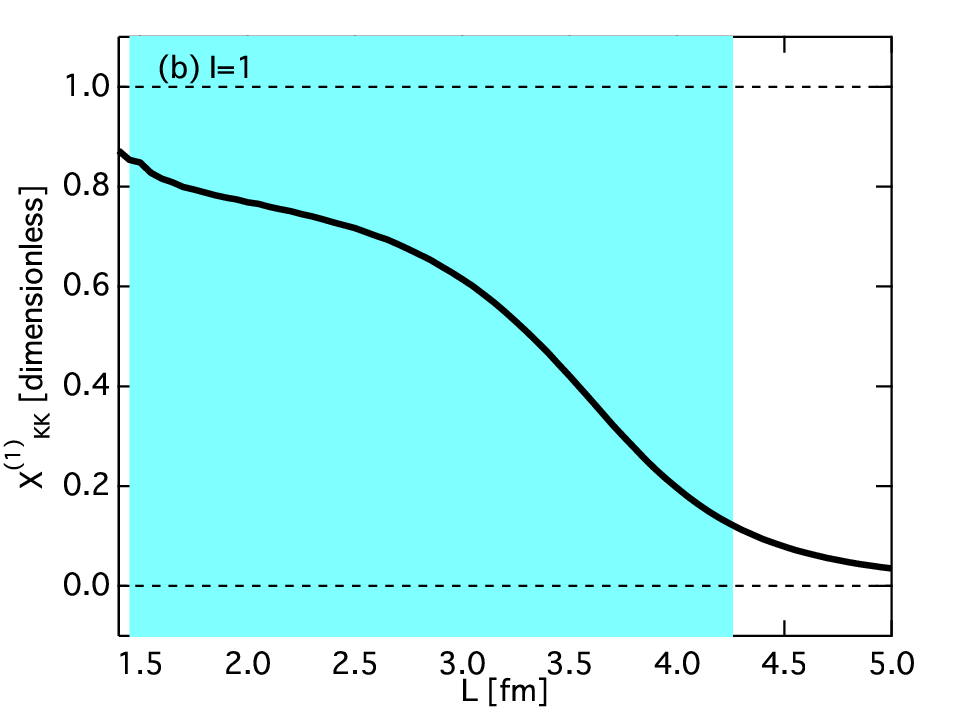}
    \includegraphics[width=8cm]{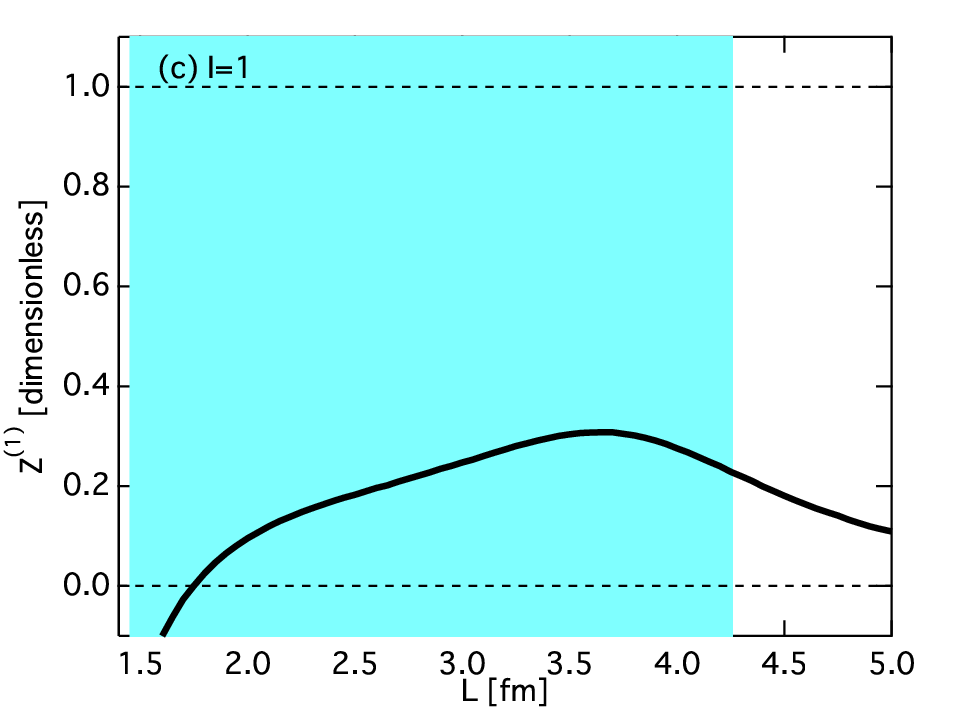}
    \caption{\label{fig:compositenessI1}
    Compositeness of the first excited state $X^{(1)}_{\pi\eta}(L)$ (a) and $X^{(1)}_{K\bar{K}}(L)$ (b) and the elementariness $Z^{(1)}(L)$ (c) in the $I=1$ scalar meson sector. The shaded areas represent the region $L_{\rm min}\leq L \leq L_{\rm max}$.}
\end{figure}

We evaluate the compositeness of $f_{0}(980)$ and $a_{0}(980)$ from the finite volume eigenstates. We choose the first excited state in the $I=1$ sector as discussed in Sec.~\ref{subsec:prescription}. On the other hand, in the $I=0$ sector, the first excited state does not exhibit the plateau corresponding to $f_{0}(980)$, because the noninteracting eigenenergy of the $\pi\pi$ scattering reaches 980 MeV around $\sim$2.7 fm. We thus choose the plateau region of the second excited state to evaluate the compositeness of $f_{0}(980)$. The $L$ dependence of the compositeness of the second excited state in the $I=0$ channel $X_{i}^{(2)}(L)$ and $Z^{(2)}(L)$ are shown in Fig.~\ref{fig:compositenessI0}, and those of the first excited state in the $I=1$ channel in Fig.~\ref{fig:compositenessI1}. In both cases, we observe that the property of the eigenstate is dominated by the channel 1 component ($\pi\pi$ in $I=0$ and $\pi\eta$ in $I=1$) at large $L$, because the eigenenergy eventually follows the scaling of the $\pi\pi$/$\pi\eta$ scattering state. We note that the elementariness $Z^{(1)}(L)$ in the $I=1$ sector becomes negative at small $L$ region. The negative value of the elementariness, even for a stable bound state, is known to occur with the energy-dependent interaction $V_{ij}(W)$~\cite{Sekihara:2016xnq}, because of the emergence of the negative norm states~\cite{Miyahara}.

%--figure---------------------------------
%--figure---------------------------------

%--figure---------------------------------
%--figure---------------------------------

With the imaginary part of the $\bar{K}K$ scattering amplitude in Fig.~\ref{fig:amplitude_scalar}, we determine $(W_{\rm min},W_{\rm max})=(953.9 ,990.7 \text{ MeV})$ for $I=0$ and $(W_{\rm min},W_{\rm max})=(913.5,1000.2 \text{ MeV})$ for $I=1$. As seen in Fig.~\ref{fig:amplitude_scalar}, the width of $a_{0}(980)$ is broader than $f_{0}(980)$, and therefore $a_{0}(980)$ has a larger energy region than that of $f_{0}(980)$. These energy regions correspond to the regions of $L$ as $(L_{\rm min},L_{\rm max})=( 1.93,2.11 \text{ fm})$ for $I=0$ and $(L_{\rm min},L_{\rm max})=( 1.45,4.25 \text{ fm})$ for $I=1$. These regions are shown by the shaded areas in Figs.~\ref{fig:compositenessI0} and \ref{fig:compositenessI1}. Averaging over these regions, we obtain the compositeness $X_{{\rm res},i}$ and the elementariness $Z_{\rm res}$ of $f_{0}(980)$ and $a_{0}(980)$ as summarized in Tables~\ref{tbl:compositenessI0} and \ref{tbl:compositenessI1}. In both cases, the $\bar{K}K$ component dominates roughly half of the wave function, i.e., 46\% in $f_{0}(980)$ and 57\% in $a_{0}(980)$.

\begin{table}[tbp]
	\begin{center}
\caption{Compositeness of the $f_{0}(980)$ resonance $X_{\text{res},i}$ and the elementariness $Z_{\rm res}$ obtained in this work. For comparison, we show $\tilde{X}_{i}$ and $\tilde{Z}$ defined in Eq.~\eqref{eq:Xtildemulti} evaluated at the pole position.}
	\begin{ruledtabular}
\begin{tabular}{lcc}
channel & This work & Residue at $W_{\rm res}$  \\
\hline
$\pi\pi$   & 0.36 & 0.02 \\ 
$\bar{K}K$ & 0.46 & 0.72 \\
others     & 0.18 & 0.26 \\ 
\end{tabular}
\label{tbl:compositenessI0}
    \end{ruledtabular}
	\end{center}
\end{table}%

\begin{table}[tbp]
	\begin{center}
\caption{Compositeness of the $a_{0}(980)$ resonance $X_{\text{res},i}$ and the elementariness $Z_{\rm res}$ obtained in this work. For comparison, we show $\tilde{X}_{i}$ and $\tilde{Z}$ defined in Eq.~\eqref{eq:Xtildemulti} evaluated at the pole position.}
	\begin{ruledtabular}
\begin{tabular}{lcc}
channel & This work & Residue at $W_{\rm res}$  \\
\hline
$\pi\eta$  & 0.24 & 0.09 \\ 
$\bar{K}K$ & 0.57 & 0.37 \\
others     & 0.19 & 0.54 \\ 
\end{tabular}
\label{tbl:compositenessI1}
    \end{ruledtabular}
	\end{center}
\end{table}%

% complex compositeness at each pole
For comparison, we also evaluate the compositeness at the pole position in infinite volume. For $f_{0}(980)$, we obtain $X_{\pi\pi}=0.01 + 0.01i$, $X_{\bar{K}K}=0.74-0.11i$, and $Z=0.25 + 0.10i$. For $a_{0}(980)$, we obtain $X_{\pi\eta}=-0.06 + 0.10i$, $X_{\bar{K}K}=0.38-0.29i$, and $Z=0.68 + 0.18i$. To interpret the complex-valued compositeness, we follow Ref.~\cite{Sekihara:2015gvw} to define real-valued quantities
\begin{align}
    \tilde{X}_{i}
    &= 
    \frac{|X_{i}|}{U+1}, \quad
    \tilde{Z}
    = 
    \frac{|Z|}{U+1}, 
    \label{eq:Xtildemulti}\\
    U
    &= \sum_{i}|X_{i}|+|Z| - 1,
    \label{eq:Udef}
\end{align}
which is the generalization of the single-channel version proposed in Refs.~\cite{Kamiya:2015aea,Kamiya:2016oao}. The results of $\tilde{X}_{i}$ and $\tilde{Z}$ are summarized in Tables~\ref{tbl:compositenessI0} and \ref{tbl:compositenessI1}. The results of $f_{0}(980)$ show a similar tendency with this work, the dominance of the $\bar{K}K$ component. The value of the $\bar{K}K$ compositeness of $f_{0}(980)$ also reasonably agrees with other estimation in Ref.~\cite{Guo:2015daa} $X^{R}_{\bar{K}K}=0.65^{+0.27}_{-0.26}$. On the other hand, for $a_{0}(980)$, the evaluation by the residue of the pole shows the dominance of the elementary component, in contrast to the $\bar{K}K$ dominance of the finite volume method. This discrepancy may partly be caused by the broader width of the $a_{0}(980)$ than $f_{0}(980)$. It may also reflect the unclear nature of the $a_{0}(980)$ pole; the determination of the pole position of $a_{0}(980)$ is still controversial~\cite{Kamiya:2015aea,Kamiya:2016oao}, and some analysis claims that $a_{0}(980)$ is not a state but a threshold cusp phenomena.

%------------------------------
\subsection{$\Lambda(1405)$ in infinite volume}\label{subsec:L1405}

% intro
The properties of the $\Lambda(1405)$ resonance have been successfully reproduced in chiral SU(3) dynamics~\cite{Kaiser:1995eg,Oset:1998it,Oller:2000fj,Jido:2003cb,Hyodo:2007jq,Hyodo:2011ur}, where the interaction kernel derived in chiral perturbation theory is iterated in the scattering equation to obtain the coupled-channel meson-baryon scattering amplitude with strangeness $S=-1$ and isospin $I=0$. The subtraction constants in the loop functions are determined by the total cross sections of the $K^{-}p$ elastic and inelastic scatterings and the threshold branching ratios. In addition, the $K^{-}p$ scattering length has recently been determined by the measurement of the kaonic hydrogen by SIDDHARTA~\cite{Bazzi:2011zj,Bazzi:2012eq,Meissner:2004jr}. Thanks to the accurate constraints by SIDDHARTA, it is now possible to discuss $\Lambda(1405)$ at the quantitative level. In fact, the meson-baryon scattering amplitude is constructed by achieving $\chi^{2}$ per degree of freedom $\sim$1 with all experimental data~\cite{Ikeda:2011pi,Ikeda:2012au}.

% ETW model
In this work, we employ the effective Tomozawa-Weinberg (ETW) model introduced in Ref.~\cite{Ikeda:2012au} for the description of the $\Lambda(1405)$ resonance. This is a simple two-channel ($\bar{K}N$ and $\pi\Sigma$) model with the leading order chiral interaction, but reasonably well reproduces the results of the full next-to-leading order (NLO) calculation including the SIDDHARTA result. The interaction kernel to be used in Eq.~\eqref{eq:amplitude} is given by
\begin{align}
    V_{ij}(W)
    &= 
    -\frac{C_{ij}}{8f_{i}f_{j}}N_{i}N_{j}
    (2W-M_{i}-M_{j}) ,
    \label{eq:WTinteraction}
\end{align}
where $N_{i}=\sqrt{E_{i}+M_{i}}$, $E_{i}=(W^{2}-m_{i}^{2}+M_{i}^{2})/2W$ is the energy of the baryon in channel $i$, $f_{i}$ is the decay constant of the meson in channel $i$, and $(m_{i},M_{i})$ are the masses of the meson and baryon in channel $i$, respectively. For the $\bar{K}N(i=1)$ and $\pi\Sigma(i=2)$ channels, the coupling strengths are given by
\begin{align}
    C_{ij}
    &= 
    2\begin{pmatrix}
    3 & -\sqrt{3/2} \\
    -\sqrt{3/2} & 4
    \end{pmatrix} .
\end{align}
We use $(m_{1},M_{1})=(495.6, 938.9\text{ MeV})$, $(m_{2},M_{2})=(138.0 , 1190.5\text{ MeV})$, $f_{1}=109.0$ MeV, $f_{2}=92.4$ MeV, $\mu_{\rm reg}= 1000$ MeV, $a_{1}(\mu_{\rm reg})=-1.79\times 10^{-3}\times 16\pi^{2}-1$, and $a_{2}(\mu_{\rm reg})=1.81\times 10^{-3}\times 16\pi^{2}-1$.\footnote{In the original paper~\cite{Ikeda:2012au}, the ETW model was constructed with physical hadron masses with isospin symmetry breaking effect. Here we use the masses in the isospin symmetric limit, which is sufficient for the present purpose.}

\begin{figure}[tbp]
    \centering
    \includegraphics[width=8cm]{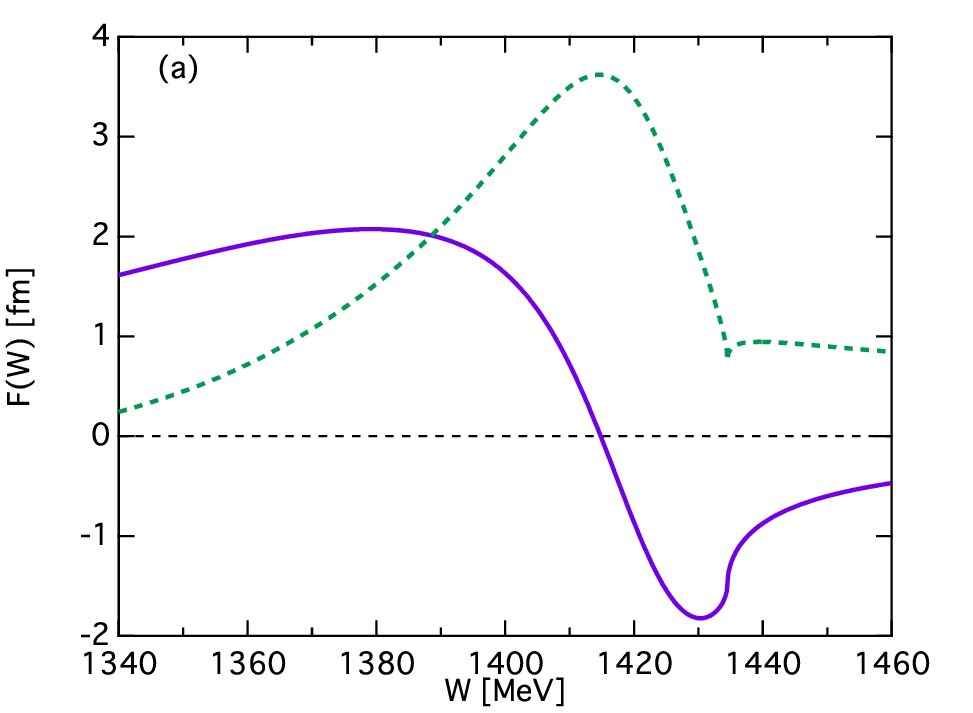}
    \includegraphics[width=8cm]{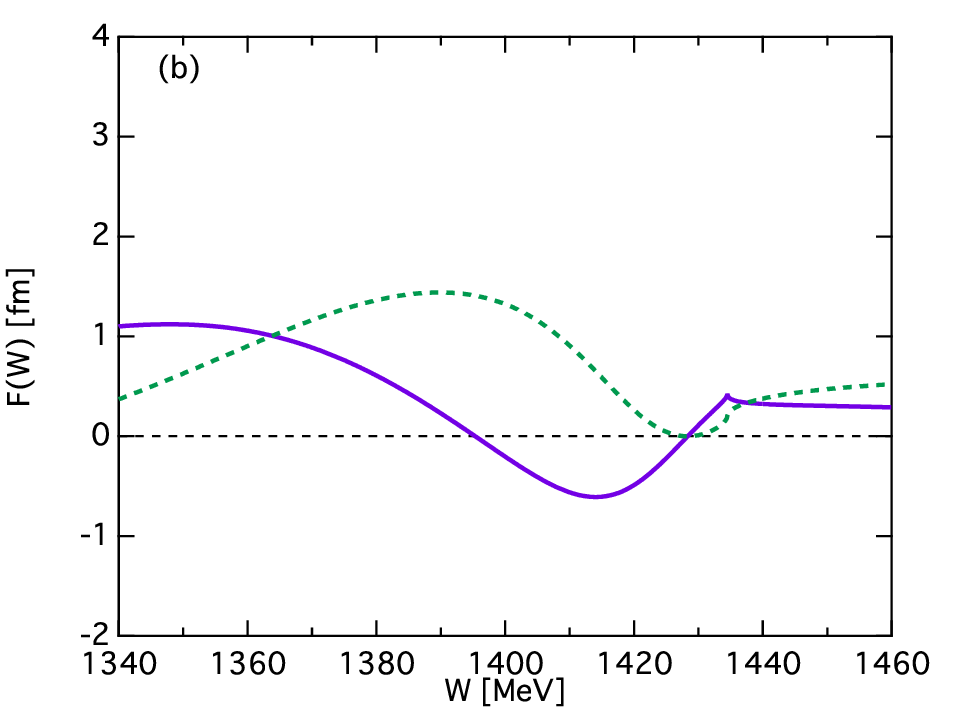}
    \caption{\label{fig:amplitude}
    Real parts (solid lines) and imaginary parts (dashed lines) of the elastic scattering amplitudes $F_{\bar{K}N}(W)$ (a) and $F_{\pi\Sigma}(W)$ (b).}
\end{figure}

% scattering amplitude
We calculate the elastic scattering amplitudes $F_{\bar{K}N}(W) =-T_{11}(W)/(8\pi W)$ and $F_{\pi\Sigma}(W) =-T_{22}(W)/(8\pi W)$ as shown in Fig.~\ref{fig:amplitude}. We observe a resonance behavior corresponding to $\Lambda(1405)$ in each amplitude. The scattering length of the $\bar{K}N$ channel is obtained as $a^{I=0}_{\bar{K}N}=-F_{\bar{K}N}(m_{1}+M_{1})=1.485-0.755i$ fm, in good agreement with the NLO result~\cite{Kamiya:2015aea}. By analytically continuing the scattering amplitude in the complex energy plane, we find two poles in the most adjacent Riemann sheet to the real energy axis between the $\bar{K}N$ and $\pi\Sigma$ thresholds. The positions of poles are found to be
\begin{align}
    W_{\rm res}^{\rm I}
    &= 1423.3-21.7i\text{ MeV} ,\\
    W_{\rm res}^{\rm II}
    &= 1371.2-65.3i\text{ MeV} ,
\end{align}
which are consistent with the results of the NLO amplitude in Refs.~\cite{Ikeda:2011pi,Ikeda:2012au}. In this way, we have obtained a reliable scattering model which successfully reproduces the experiential data. We see that two complex eigenstates are associated with one resonance structure~\cite{Jido:2003cb}.

%--figure---------------------------------
%--figure---------------------------------

%------------------------------
\subsection{$\Lambda(1405)$ in finite volume}\label{subsec:L1405FV}

\begin{figure}[tbp]
    \centering
    \includegraphics[width=8cm]{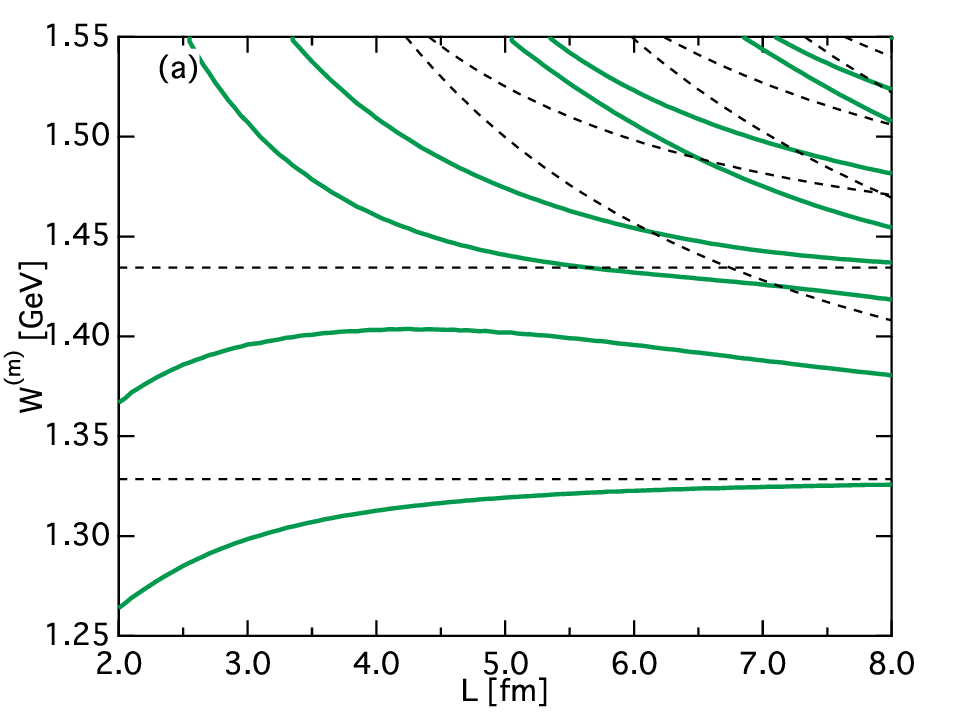}
    \includegraphics[width=8cm]{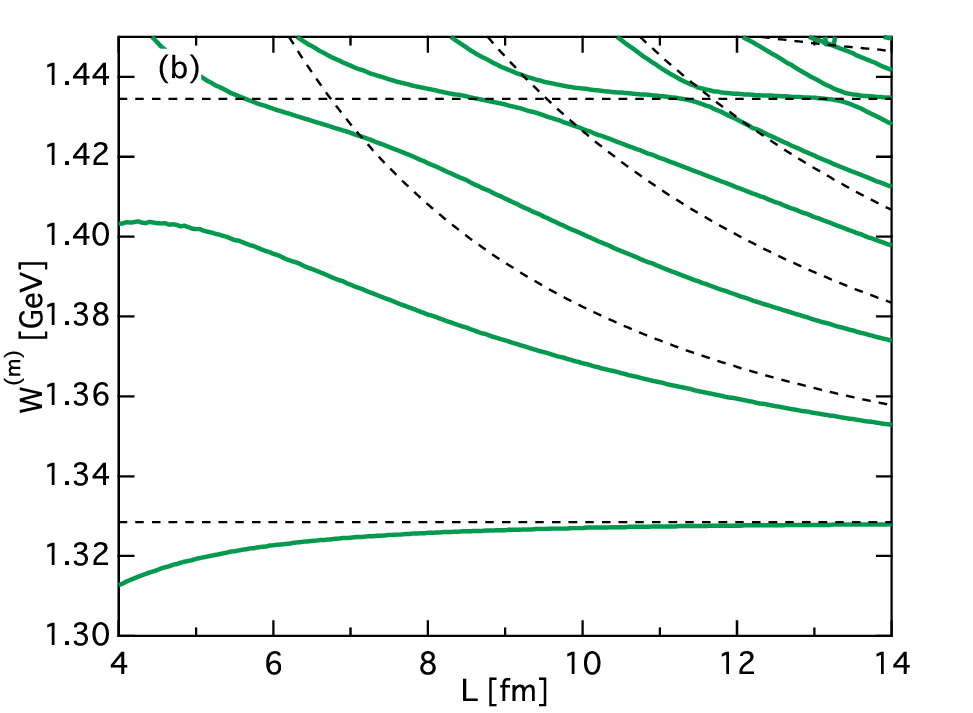}
    \caption{\label{fig:spectrum}
    Eigenenergies $W^{(m)}(L)$ in finite volume in the $\bar{K}N$-$\pi\Sigma$ sector (solid lines) in 2 fm $<L<8$ fm (a) and 4 fm $<L<14$ fm (b). The noninteracting eigenenergies $W_{\bm{n},1}(L)$ and $W_{\bm{n},2}(L)$ are shown by dashed lines for comparison.}
\end{figure}

As in Sec.~\ref{subsec:scalarFV}, we put this model in a box of size $L$. The eigenenergies $W^{(m)}(L)$ of the ground state ($m=0$) to the seventh excited state ($m=7$) are shown in Fig.~\ref{fig:spectrum} as functions of the system size $L$, together with the noninteracting eigenenergies (dashed lines). In Fig.~\ref{fig:spectrum}(a), the energy of the first excited state $W^{(1)}(L)$ shows a flat $L$ dependence around 1.4 GeV, indicating the existence of a resonance. The qualitative feature of the energy levels is consistent with the energy levels found in other models (see Fig.~3 of Ref.~\cite{Lage:2009zv}, Fig.~3 of Ref.~\cite{Doring:2011ip}, and Fig.~1 in Ref.~\cite{MartinezTorres:2012yi}). At larger $L$ [Fig.~\ref{fig:spectrum}(b)], the noninteracting $\pi\Sigma$ scattering energies $W_{\bm{n},2}$ cross 1400 MeV, and the full eigenenergies $W^{(m)}(L)$ show avoided level crossings, although the signature is not very obvious. These behaviors indicate the existence of \textit{one} energy level corresponding to $\Lambda(1405)$. Although the finite-volume energy levels were studied in Refs.~\cite{Doring:2011ip,MartinezTorres:2012yi,Molina:2015uqp}, the relation between the number of complex poles and that of the finite-volume energy levels was not clarified. As we show in Appendix~\ref{appendix:FVstates}, the number of the discrete eigenstates in finite volume is determined by the number of $\pi/2$ crossing of the phase shift $\delta$. Because the scattering amplitude is related to the phase shift as $F=(e^{2i\delta}-1)/(2ik)$, $\delta=\pi/2$ corresponds to $\re F=0$ and $\im F=1/k\neq 0$. In Fig.~\ref{fig:amplitude}, this occurs only once in between the $\pi\Sigma$ and $\bar{K}N$ thresholds.\footnote{The real part of the $\pi\Sigma$ amplitude crosses zero again around $W=1430$ MeV, but the imaginary part also vanishes simultaneously. This corresponds to $\delta=0$ which is not the signal of a resonance. However, the appearance of $\delta=0$ has a different significance; for instance, it is related to the Castillejo-Dalitz-Dyson pole~\cite{Castillejo:1956ed,Kamiya:2016oao} and the Ramsauer-Townsend effect~\cite{Yamaguchi:2016kxa}.} Thus, there is one finite-volume eigenstate which represents $\Lambda(1405)$, in spite of two complex poles in infinite volume. In other words, the single finite-volume eigenstate represents \textit{both} the resonance eigenstates in the infinite volume.

%--figure---------------------------------
%--figure---------------------------------

\begin{figure}[tbp]
    \centering
    \includegraphics[width=8cm]{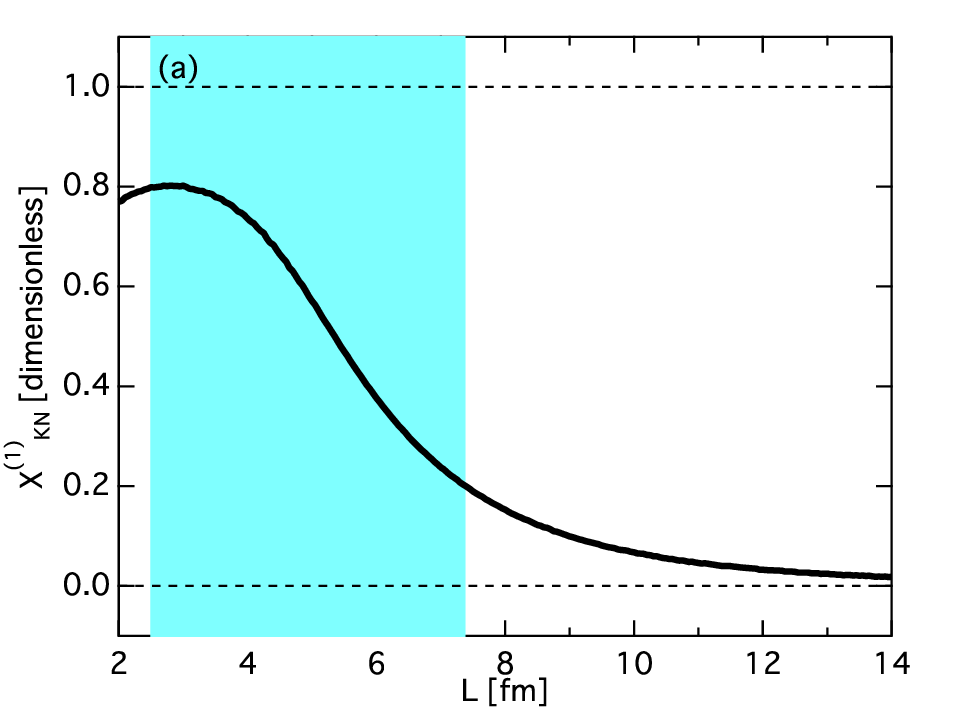}
    \includegraphics[width=8cm]{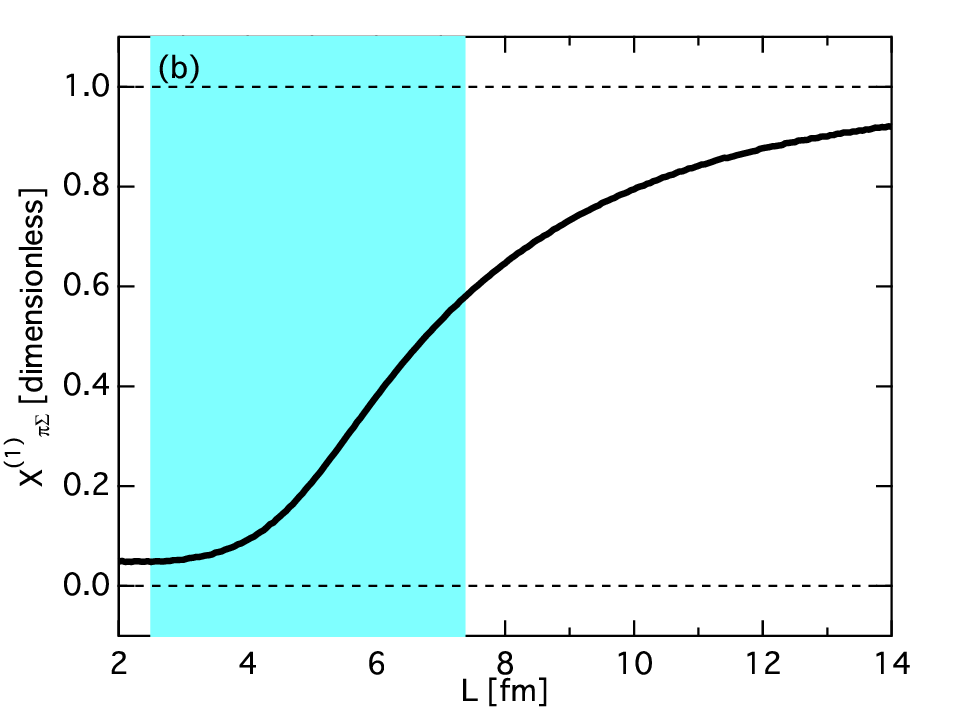}
    \includegraphics[width=8cm]{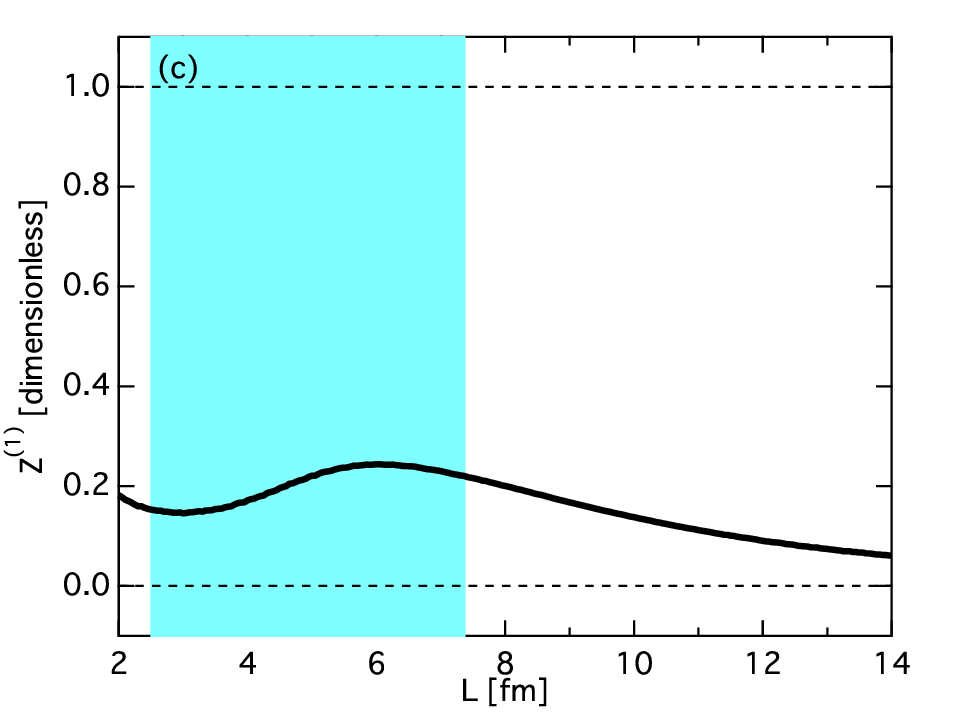}
    \caption{\label{fig:compositeness}
    Compositeness of the first excited state $X^{(1)}_{\bar{K}N}(L)$ (a) and $X^{(1)}_{\pi\Sigma}(L)$ (b) and the elementariness $Z^{(1)}(L)$ (c). The shaded areas represent the region $L_{\rm min}\leq L \leq L_{\rm max}$.}
\end{figure}

% results
We determine the compositeness of $\Lambda(1405)$ from the first excited state in finite volume. In Fig.~\ref{fig:compositeness}, we show $X^{(1)}_{\bar{K}N}(L)$, $X^{(1)}_{\pi\Sigma}(L)$, and $Z^{(1)}(L)$ as functions of $L$. We first determine the resonance energy from the imaginary part of the $\bar{K}N$ scattering amplitude, which leads to $(W_{\rm min}$, $W_{\rm max})=(1385.2,1430.2\text{ MeV})$. The corresponding region of $L$ for the first excited state is found to be $(L_{\rm min}$, $L_{\rm max})=(2.48,7.35\text{ fm})$. Averaging $X^{(1)}_{i}(L)$ and $Z^{1}(L)$ over this region, we obtain the compositeness and elementariness shown in Table~\ref{tbl:compositeness}. The $\bar{K}N$ channel occupies 58\% of the wave function and is the main component of $\Lambda(1405)$. At the same time, $\pi\Sigma$ channel (26\%) and other components (16\%) are also necessary to form $\Lambda(1405)$.\footnote{The contribution from the two-body channels which are not included in the model space (such as $\eta\Lambda$) is represented by $Z_{\rm res}$~\cite{Sekihara:2014kya}.} We note that the values of $X^{(1)}_{\bar{K}N}(L)$, $X^{(1)}_{\pi\Sigma}(L)$ gradually change within the region $(L_{\rm min}<L<L_{\rm max})$ as shown in Fig.~\ref{fig:compositeness}. This can be interpreted as a consequence of the existence of two complex poles with different nature in infinite volume. This point can be further clarified below.  

% complex compositeness at each pole
For comparison, we also evaluate the compositeness at the pole position in infinite volume. At the higher energy pole position $W_{\rm res}^{\rm I}$, we obtain $X_{\bar{K}N}= 0.97+0.10i$, $X_{\pi\Sigma}=-0.03-0.20i$, and $Z=0.07+0.10i$. The results at $W_{\rm res}^{\rm II}$ are $X_{\bar{K}N}=-0.20-0.17i$, $X_{\pi\Sigma}=0.42+0.58i$, and $Z=0.78-0.41i$. The results of $\tilde{X}_{i}$ and $\tilde{Z}$ by Eq.~\eqref{eq:Xtildemulti} are summarized in Table~\ref{tbl:compositeness}. The value of the $\bar{K}N$ compositeness at $W_{\rm res}^{\rm I}$ reasonably agrees with other studies; $\tilde{X}_{\bar{K}N}=1.0^{+0.0}_{-0.4}$~\cite{Kamiya:2016oao} and $X^{R}_{\bar{K}N}=0.82^{+0.36}_{-0.17}$~\cite{Guo:2015daa}. The eigenstate at $W_{\rm res}^{\rm I}$ is dominated by the $\bar{K}N$ channel about 80 \%, while the state at $W_{\rm res}^{\rm II}$ consists of three components with comparable magnitudes. The comparison of the results in Table~\ref{tbl:compositeness} suggests that the compositeness determined from the finite-volume eigenstate includes the contribution from both poles. This is in accordance with the fact that the finite-volume eigenenergy represents the two complex poles in infinite volume. Quantitatively, $W_{\rm res}^{\rm I}$ has a larger contribution to $X_{\rm res}$, presumably because it is closer to the real axis than $W_{\rm res}^{\rm II}$.

% pi Sigma version
The peak structure of the $\bar{K}N$ amplitude is not identical with that in the $\pi\Sigma$ amplitude, because of the double-pole nature of $\Lambda(1405)$~\cite{Jido:2003cb}. If we determine the resonance energy by the $\pi\Sigma$ amplitude, we find $(W_{\rm min}$, $W_{\rm max})=(1353.4,1413.0\text{ MeV})$ and $(L_{\rm min}$, $L_{\rm max})=(1.79,13.79\text{ fm})$. The resonance energy region is lower than that in the $\bar{K}N$ amplitude, and the range of $L$ is increased. By calculating the compositeness, we obtain $X_{\text{res},\bar{K}N}=0.31$, $X_{\text{res},\pi\Sigma}=0.53$, and $Z_{\text{res}}=0.16$. Namely, the relative importance of the $\pi\Sigma$ component increases. It is natural to expect that the $\pi\Sigma$ amplitude puts more weight on the pole $W_{\rm res}^{\rm II}$ which has a larger $\pi\Sigma$ compositeness $\tilde{X}_{\pi\Sigma}$. In other words, the channel dependence of the resonance energy is an indication of the double-pole nature of $\Lambda(1405)$. Note however that the imaginary part of the pole position $W_{\rm res}^{\rm II}$ is not small, and the quantitative interpretation of the structure is somewhat ambiguous for the $W_{\rm res}^{\rm II}$ pole, because of the larger $U$ in Eq.~\eqref{eq:Udef}.

%--figure---------------------------------
%--figure---------------------------------

\begin{table}[tbp]
	\begin{center}
\caption{Compositeness of the $\Lambda(1405)$ resonance $X_{\text{res},i}$ and the elementariness $Z_{\rm res}$ obtained in this work. For comparison, we show $\tilde{X}_{i}$ and $\tilde{Z}$ defined in Eq.~\eqref{eq:Xtildemulti} evaluated at each pole position.}
	\begin{ruledtabular}
\begin{tabular}{lccc}
Channel & This work & Residue at $W_{\rm res}^{\rm I}$ & Residue at $W_{\rm res}^{\rm II}$  \\
\hline
$\bar{K}N$  & 0.56 & $0.75$ & $0.14$ \\ 
$\pi\Sigma$ & 0.24 & $0.15$ & $0.39$ \\
others      & 0.20 & $0.10$ & $0.47$ \\ 
\end{tabular}
\label{tbl:compositeness}
    \end{ruledtabular}
	\end{center}
\end{table}%

%%%%%%%%%%%%%%%%%%%%%%%%%%%%%%%%%%%%%%%%%%%%%%%%%%%%%%%%%%%%%%%%%%%%%%%%
\section{Summary}\label{sec:summary}
%%%%%%%%%%%%%%%%%%%%%%%%%%%%%%%%%%%%%%%%%%%%%%%%%%%%%%%%%%%%%%%%%%%%%%%%

In this paper, the composite structure of hadron resonances is discussed by using the finite-volume system. It is shown that all the eigenstates in finite volume have a well-defined compositeness which can be interpreted as a probability. By identifying the finite-volume eigenstate which reflects the property of the resonance from the volume dependence of the eigenenergy, we define the compositeness of the resonance. In single-channel scattering models, we show that our prescription gives results in accordance with other proposals to quantify the structure of resonances.

As an application, we study the compositeness of the scalar mesons $f_{0}(980)$ and $a_{0}(980)$ and the $\Lambda(1405)$ resonance. We find that the dominant component of $f_{0}(980)$ and $a_{0}(980)$ is the $\bar{K}K$ molecular contribution, which amounts to about half of their wave function. For the $\Lambda(1405)$ resonance, we show that a single finite-volume eigenstate is responsible, in spite of the two complex eigenstates in infinite volume. The $\bar{K}N$ component is found to be 58\%, and the $\pi\Sigma$ and other components also contribute to the structure of $\Lambda(1405)$. In contrast to the previous works which focus on the individual complex eigenstate, our results reflect the structure of $\Lambda(1405)$ as a whole, including the contribution from both poles. Given the recent works which shows that the higher energy pole $W_{\rm res}^{\rm I}$ is dominated by the $\bar{K}N$ molecular structure about $\sim$80\%~\cite{Kamiya:2016oao,Guo:2015daa}, the present results indicate the importance of the coupled-channel dynamics for the description of $\Lambda(1405)$ as discussed in Ref.~\cite{Hyodo:2007jq}.

%%%%%%%%%%%%%%%%%%%%%%%%%%%%%%%%%%%%%%%%%%%%%%%%%%%%%%%%%%%%%%%%%%%%%%%%
\section*{Acknowledgments}
%%%%%%%%%%%%%%%%%%%%%%%%%%%%%%%%%%%%%%%%%%%%%%%%%%%%%%%%%%%%%%%%%%%%%%%%

This work was supported in part by JSPS KAKENHI Grant No. JP16K17694 and by the Yukawa International Program for Quark-Hadron Sciences (YIPQS).

\appendix
%%%%%%%%%%%%%%%%%%%%%%%%%%%%%%%%%%%%%%%%%%%%%%%%%%%%%%%%%%%%%%%%%%%%%%%%
\section{Wavefunctions of resonances and eigenstates in finite size system}\label{appendix:resonancewf}
%%%%%%%%%%%%%%%%%%%%%%%%%%%%%%%%%%%%%%%%%%%%%%%%%%%%%%%%%%%%%%%%%%%%%%%%

% intro
In Sec.~\ref{subsec:prescription}, we argue that the property of a resonance in infinite volume is reflected in the finite-volume eigenstate whose eigenenergy is close to the resonance energy. Here we explicitly demonstrate this statement using a potential problem in one-dimensional quantum mechanics. We consider the Schr\"odinger equation (by setting $\hbar=1$ and the mass $M=1$),\footnote{In this unit, all quantities are measured by the dimension of length.}
\begin{align}
   \left(-\frac{1}{2}\frac{d^{2}}{dx^{2}}+V(x)\right)\psi(x)
   &= E\psi(x)
   \label{eq:Schroedinger1d} ,
\end{align}
where $E$ and $\psi(x)$ are the eigenenergy and the wave function of the eigenstate. 

% wavefunction
Here we adopt the barrier potential~\cite{Moiseyev}
\begin{align}
   V(x)
   &= \begin{cases}
   \infty & x\leq0 \\
   0 & 0< x \leq  b \\
   -V_{0} & b<x
   \end{cases} ,
   \label{eq:barrier}
\end{align}
with $V_{0}>0$.\footnote{The following discussion can be performed by using other potentials which vanish at large $x$. For instance, one can use the attractive square well potential with a finite barrier, which is more similar to the physical situation studied in the main text. It is straightforward to perform the similar analysis using the resonance eigenenergies and wave functions in Ref.~\cite{delaMadrid:2002cz}.} This is equivalent to the repulsive rectangular potential with the shift of the origin of the energy. Because the boundary condition is imposed only at $x=0$, there are scattering state solutions which form a continuum spectrum with real $E>-V_{0}$. The wavefunction of the scattering state is analytically given by
\begin{align}
   \psi(x)
   &= \begin{cases}
   \sin(kx) & 0\leq x \leq  b \\
   A(k)e^{-iq(x-b)}+B(k)e^{iq(x-b)} & b<x
   \end{cases} , 
   \label{eq:wavefunction1d} \\
   A(k)
   &=\frac{1}{2}\left(\sin (kb)+i\frac{k}{q}\cos(kb)\right) , \\
   B(k)
   &=\frac{1}{2}\left(\sin (kb)-i\frac{k}{q}\cos(kb)\right) ,
\end{align}
with $k=\sqrt{2E}$ and $q=\sqrt{2(E+V_{0})}=\sqrt{k^{2}+2V_{0}}$. Note that the wave function of the scattering state is not normalizable. Here we fix the amplitude of the wave function in the interaction region $0\leq x \leq  b$ to be unity. 

\begin{figure}[tbp]
    \centering
%    \hspace*{1cm}
    \includegraphics[width=7.5cm]{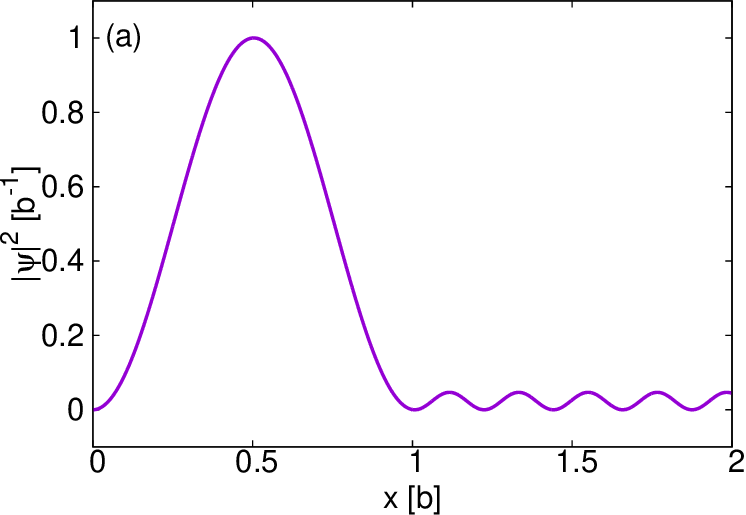}
%    \hspace*{1cm}
    \includegraphics[width=7.5cm]{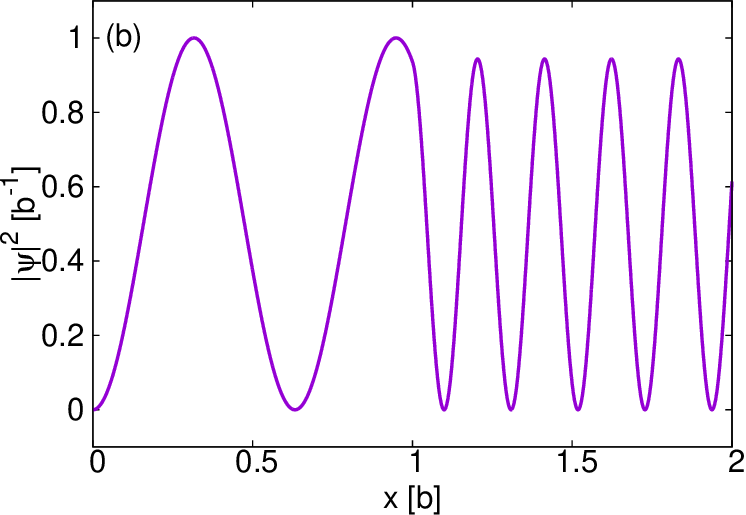}
%    \vspace{0.5cm}
    \caption{\label{fig:wf1d}
    Wave function $|\psi(x)|^{2}$ as a function of $x$ near the resonance [(a) $\tilde{E}= 0.985$] and wave function off the resonance positions [(b) $\tilde{E}= 2.500$].}
\end{figure}

% resonances
A discrete eigenenergy is given by the pole of the $S$ matrix which is defined by the ratio of the amplitude of the outgoing wave to that of the incoming wave in the asymptotic region as $S(k)=B(k)/A(k)$. There is a pole of $B(k)$ at $E=-V_{0}$, but $A(k)$ also diverges at this point and $S(k)=-1$ is finite. Thus, the discrete eigenstates are determined by the zeros of $A(k)$, namely,
\begin{align}
   \tan(k_{\rm res}b)
   &= -i\frac{k_{\rm res}}{\sqrt{k_{\rm res}^{2}+2V_{0}}} .
\end{align}
It is clear that no solution is found with a real $k_{\rm res}$. There are however solutions with complex $k_{\rm res}$ through the analytic continuation, which represent resonances. We define the dimensionless eigenenergies\footnote{The normalization of $\tilde{E}$ is chosen such that the eigenenergies in the $V_{0}\to\infty$ limit are given by $\tilde{E}^{(m)}=(m+1)^{2}$~\cite{Moiseyev}.}
\begin{align}
   \tilde{E}^{(m)}
   &= \left(\frac{k_{\rm res}b}{\pi}\right)^{2} ,
   \quad \re[\tilde{E}^{(m+1)}]> \re[\tilde{E}^{(m)}] .
\end{align}
With $V_{0}=100 b^{-2}$, the first three eigenenergies are obtained as
\begin{align}
   \tilde{E}^{(0)}
   &= 0.985-0.139i, \\
   \tilde{E}^{(1)}
   &= 3.946-0.544i, \\
   \tilde{E}^{(2)}
   &= 8.891-1.186i .
\end{align}
The resonance phenomena can be seen in the behavior of the wave function of the scattering state with a real energy. We plot the wave function $|\psi(x)|^{2}$ near the lowest resonance $\tilde{E}=(2b^{2}/\pi^{2})E=0.985= \re [\tilde{E}^{(0)}]$ in Fig.~\ref{fig:wf1d}. For comparison, we also plot the wave function at $\tilde{E}=2.500$ which is in between the two resonances. The wave function is localized in the interaction region $0\leq x\leq b$ when the energy is close to the resonance position [Fig.~\ref{fig:wf1d}(a)], and it behaves as a plane wave when the energy is away from the resonances [Fig.~\ref{fig:wf1d}(b)].

%--figure---------------------------------
%--figure---------------------------------

% finite size
Next, we consider the eigenstates in the corresponding finite system of size $L$. We impose the Dirichlet boundary condition for the wave function $\psi_{\rm FS}(L)=0$. This is equivalent to consider the potential 
\begin{align}
   V_{\rm FS}(x)
   &= \begin{cases}
   \infty & x\leq 0,\ L\leq x\\
   0 & 0< x \leq  b \\
   -V_{0} & b<x<L
   \end{cases} .
   \label{eq:barrierFV}
\end{align}
Because of two boundary conditions at $x=0$ and $x=L$, the eigenmomentum is discretized to satisfy
\begin{align}
   \tan(k_{\rm FS}b)
   &= \frac{k_{\rm FS}}{q_{\rm FS}} 
   \tan\left[q_{\rm FS}(b-L)\right]
   \label{eq:eigenmomentum1d} ,
\end{align}
where $q_{\rm FS}=\sqrt{k_{\rm FS}^{2}+2V_{0}}$. This equation has solutions with real $k_{\rm FS}$, from which the dimensionless eigenenergies are defined as
\begin{align}
   \tilde{E}_{\rm FS}^{(m)}
   &= \left(\frac{k_{\rm FS}b}{\pi}\right)^{2} ,
   \quad \tilde{E}_{\rm FS}^{(m+1)}> \tilde{E}_{\rm FS}^{(m)} .
\end{align}
Because the system size $L$ is included in Eq.~\eqref{eq:eigenmomentum1d}, the eigenenergies $\tilde{E}_{\rm FS}^{(m)}$ depend on $L$. In the noninteracting limit, the eigenmomentum should be $k_{\rm FS,nonint.}^{(m)}=\sqrt{\pi^{2} m^{2}/L^{2}-2V_{0}}$.\footnote{Because we regard $0<x\leq b$ as the interaction region, the noninteracting limit is defined by $V_{\rm FS}(x)=-V_{0}$ for $0<x\leq L$. Note the factor $2$ difference of the coefficient of $\pi m/L$ from the periodic boundary condition.} 

\begin{figure}[tbp]
    \centering
%    \hspace*{1cm}
    \includegraphics[width=7.5cm]{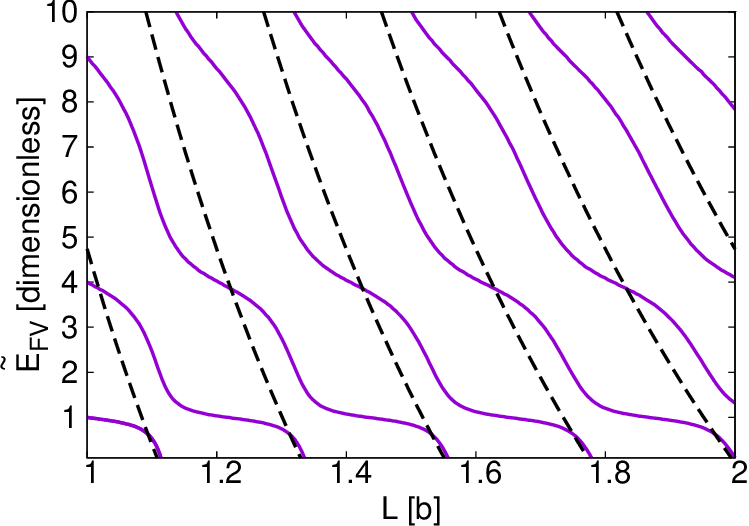}
%    \vspace{0.5cm}
    \caption{\label{fig:spectrum1d}
    Energy spectrum of the barrier potential in Eq.~\eqref{eq:barrierFV} as a function of the size $L$. Dashed lines represent the noninteracting energies.}
\end{figure}

% finite size spectrum
The discrete energy spectrum as a function of $L$ is shown in Fig.~\ref{fig:spectrum1d} with $V_{0}=100b^{-2}$. For comparison, we plot the noninteracting levels $\tilde{E}_{\rm FS,nonint.}^{(m)}=b^{2}m^{2}/L^{2}-2V_{0}b^{2}/\pi^{2}$ by the dashed lines. We observe a clear plateau of the energy levels around the lowest resonance energy $\re[\tilde{E}^{(0)}]= 0.985$, while the plateau structures of the higher resonances are not very obvious.

%--figure---------------------------------
%--figure---------------------------------

\begin{figure}[tbp]
    \centering
%    \hspace*{1cm}
    \includegraphics[width=7.5cm]{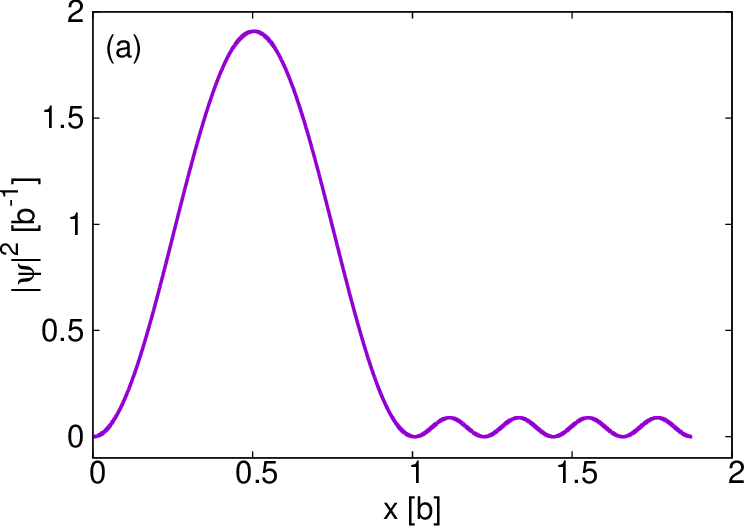}
%    \hspace*{1cm}
    \includegraphics[width=7.5cm]{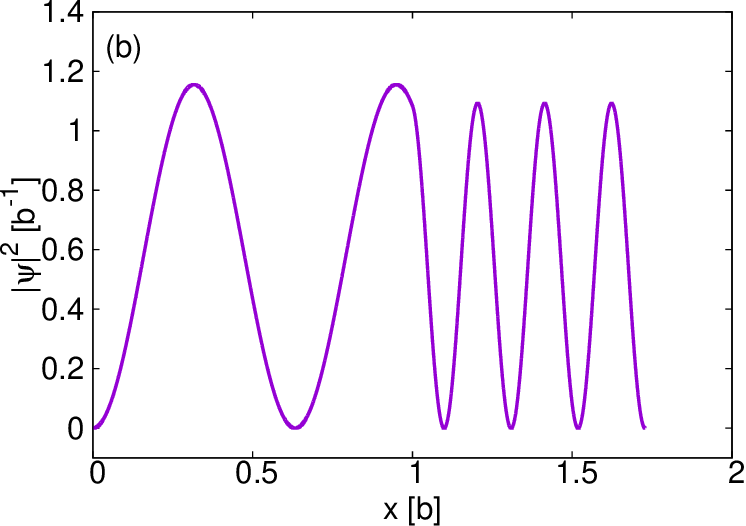}
%    \vspace{0.5cm}
    \caption{\label{fig:wf1dFS}
    Wave function $|\psi_{\rm FS}(x)|^{2}$ as a function of $x$ near the resonance [(a) $L=1.875b$, $\tilde{E}_{\rm FS}^{(5)}\simeq 0.986$] and wavefunction off the resonance positions [(b) $L=1.728b$, $\tilde{E}_{\rm FS}^{(5)}\simeq 2.500$].}
\end{figure}

% finite size wavefunction
The wave function of an eigenstate with $k_{\rm FS}$ is given by
\begin{align}
   \psi_{\rm FS}(x)
   &= \begin{cases}
   C(k_{\rm FS})\sin(k_{\rm FS}x) & 0\leq x \leq  b \\
   D(k_{\rm FS})\sin\left[q_{\rm FS}(x-L)\right] &  b<x\leq L
   \end{cases} , 
   \label{eq:wavefunction1dFV}\\
   C(k_{\rm FS})
   &=\Biggl[\frac{b}{2}-\frac{\sin(2k_{\rm FS}b)}{4k_{\rm FS}}+
   \frac{\sin^{2} (k_{\rm FS}b)}{\sin^{2}\left[q_{\rm FS}(b-L)\right]} \nonumber \\
   &\quad \times
   \Biggl(-\frac{b-L}{2}+\frac{\sin \left[2q_{\rm FS}(b-L)\right]}{4q_{\rm FS}}
   \Biggr)\Biggr]^{-1/2} ,\\
   D(k_{\rm FS})
   &=\frac{\sin (k_{\rm FS}b)}{\sin\left[q_{\rm FS}(b-L)\right]}C(k) ,
\end{align}
which is normalized as 
\begin{align}
   \int_{0}^{L}|\psi_{\rm FS}(x)|^{2} dx
   &=1 .
\end{align}
In Fig.~\ref{fig:wf1dFS}, we compare the wave function $|\psi_{\rm FS}(x)|^{2}$ near the resonance ($L=1.875b$, $\tilde{E}_{\rm FS}^{(5)}\simeq 0.986$) and that off the resonance ($L=1.728b$, $\tilde{E}_{\rm FS}^{(5)}\simeq 2.500$). As in the infinite system, the wave function is localized when the eigenenergy is close to the resonance energy.

%--figure---------------------------------
%--figure---------------------------------

% localization parameter
Now we are in a position to discuss the consequence of the resonance phenomena in the eigenstates in the finite system. From Figs.~\ref{fig:wf1d} and \ref{fig:wf1dFS}, we see that the resonance phenomena is characterized by the localization of the wave function in the interaction region. To quantify this property, we define the localization parameter $R$ as the ratio of the squared amplitude of the wave function in the interaction region ($0\leq x\leq b$) to the outside region ($b\leq x$). From the wave function in the infinite system $\psi(x)$ in Eq.~\eqref{eq:wavefunction1d}, the localization parameter is given by
\begin{align}
   R(k)
   &=\frac{1}{4|A(k)|^{2}}
   =\frac{1}{\sin^{2}(kb)+\frac{k^{2}}{q^{2}}\cos^{2}(kb)} .
\end{align}
The wavefunction in the finite system $\psi_{\rm FS}(x)$ in Eq.~\eqref{eq:wavefunction1dFV} gives
\begin{align}
   R_{\rm FS}(k_{\rm FS})
   &=\left|\frac{C(k_{\rm FS})}{D(k_{\rm FS})}\right|^{2}
   =\frac{\sin^{2}\left[q_{\rm FS}(b-L)\right]}{\sin^{2}(k_{\rm FS}b)} .
\end{align}
If the wave function is well localized in the interaction region, the value of $R$ becomes large. On the other hand, if the wave function behaves as a scattering state, we obtain $R\sim 1$.

% resonance momentum
To characterize the resonance structure in the infinite system, we define the real-valued resonance momentum
\begin{align}
   k^{(m)}
   &\equiv\frac{\pi}{b}\sqrt{\re [\tilde{E}^{(m)}]}.
\end{align}
The localization parameters of the three lowest resonances with $V_{0}=100b^{-2}$ is calculated as
\begin{align}
   R(k^{(0)})
   &= 21.33  , \\
   R(k^{(1)})
   &= 6.079  , \\
   R(k^{(2)})
   &= 3.254 .
\end{align}
This result shows that the localization is prominent when the width of the resonance is narrow. This explains the behavior of the energy spectrum in Fig.~\ref{fig:spectrum1d} where the eigenenergy in the finite system shows a weak $L$ dependence near the narrow resonance. Because the wave function is well localized near the narrow resonance, the energy levels are less affected by the modification of the boundary.

\begin{figure}[tbp]
    \centering
%    \hspace*{1cm}
    \includegraphics[width=7.5cm]{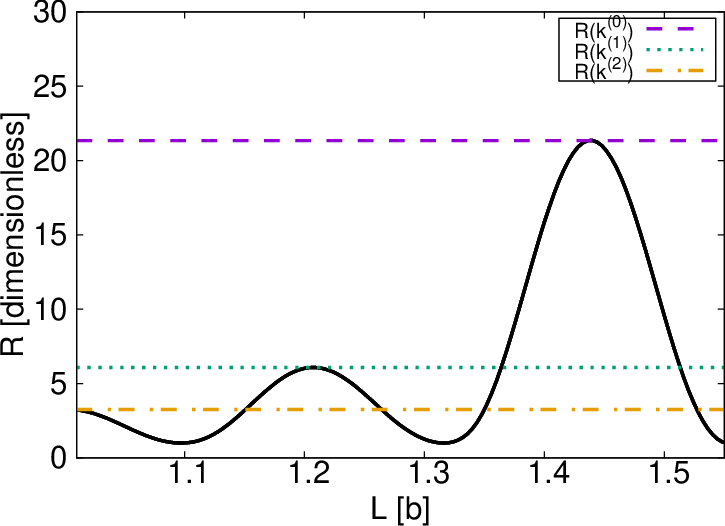}
%    \vspace*{0.5cm}
    \caption{\label{fig:Rparam}
    Localization parameter $R_{\rm FS}$ of the $m=2$ state in the finite system. Horizontal lines represent the localization parameters of the first three resonance states in the infinite system.}
\end{figure}

% comparison
In Fig.~\ref{fig:Rparam}, we plot $R_{\rm FS}$ of the $m=2$ state as a function of the system size $L$. We see that the value of $R_{\rm FS}$ is close to $R$ when the eigenenergy in the finite system approaches the resonance energy ($L\sim 1.455 b$ for $m=0$, $L\sim 1.210 b$ for $m=1$ and $L\sim 1.006 b$ for $m=2$). This feature is observed not only in the narrow state $(m=0)$ but also in the broader states $(m=1,2)$. In fact, it follows from Eq.~\eqref{eq:eigenmomentum1d} that the ratio of the localization parameters becomes unity when $k_{\rm FS}\to k^{(m)}$:
\begin{align}
   &\quad 
   \left.\frac{R_{\rm FS}(k_{\rm FS})}{R(k^{(m)})}
   \right|_{k_{\rm FS}\to k^{(m)}} \nonumber \\
   &=\left.\sin^{2}\left[q_{\rm FS}(b-L)\right]
   \left[1+\frac{k^{(m) 2}}{q^{(m) 2}}\cot^{2}(k^{(m)}b)\right]
   \right|_{k_{\rm FS}\to k^{(m)}} \\
   &= \sin^{2}\left[q^{(m)}(b-L)\right]
   \left[1+\cot^{2}[q^{(m)}(b-L)]\right] \\
   &=1 ,
\end{align}
with $q^{(m)}=\sqrt{k^{(m)2}+2V_{0}}$. In this way, we find that the characteristic localization of the resonance wave function is well reflected in the finite-system eigenstate which has a similar energy with $\re [\tilde{E}^{(m)}]$.

%--figure---------------------------------
%--figure---------------------------------

%%%%%%%%%%%%%%%%%%%%%%%%%%%%%%%%%%%%%%%%%%%%%%%%%%%%%%%%%%%%%%%%%%%%%%%%
\section{Number of eigenstates in finite volume}\label{appendix:FVstates}
%%%%%%%%%%%%%%%%%%%%%%%%%%%%%%%%%%%%%%%%%%%%%%%%%%%%%%%%%%%%%%%%%%%%%%%%

In Sec.~\ref{subsec:L1405FV}, we have seen that there is a single finite-volume eigenstate which represents the $\Lambda(1405)$ resonance, although the number of the complex eigenstates in infinite volume is 2. In this Appendix, we show that the number of the resonance eigenstates in finite volume is determined by the behavior of the phase shift on the real energy axis, rather than the poles in the complex energy plane. 

% Luescher formula
We use L\"uscher's formula which relates the finite-volume eigenenergies $E_{\rm FV}$ with the infinite-volume phase shift $\delta(E)$ as~\cite{Luscher:1986pf,Luscher:1990ux,Beane:2003da} 
\begin{align}
   \sqrt{2\mu E_{\rm FV}}\cot\delta(E_{\rm FV})
   &= -\frac{2}{\mu }I_{\rm FV}(E_{\rm FV}) ,
   \label{eq:Luescher} \\
   I_{\rm FV}(E)
   &= \frac{1}{L^{3}}
   \sum_{\bm{n}}\frac{1}{E-E_{\bm{n}}},
   \label{eq:IFVapp}
\end{align}
where $\mu$ is the reduced mass of the system and $E_{\bm{n}}=2\pi^{2}|\bm{n}|^{2}/(\mu L^{2})$ is the noninteracting eigenenergy. Any eigenenergy $E_{\rm FV}$ in finite volume satisfies this equation. In the lattice QCD simulation, this formula is used to determine the infinite-volume phase shift from the QCD eigenenergies measured in finite volume. Here we use it to study the finite-volume eigenenergy $E_{\rm FV}$ for a given phase shift $\delta(E)$. For later convenience, we choose the range of the phase shift as $-\pi/2< \delta\leq \pi/2$, so that the sign of $\delta$ coincides with that of $\cot\delta$.

% I properties
Before we get started, let us recall the properties of the function $I_{\rm FV}(E)$. Taking the energy derivative of Eq.~\eqref{eq:IFVapp}, we obtain
\begin{align}
   I^{\prime}_{\rm FV}(E)
   &= -\frac{1}{L^{3}}
   \sum_{\bm{n}}\frac{1}{(E-E_{\bm{n}})^{2}}<0,
   \label{eq:IFVprime}
\end{align}
namely, $I_{\rm FV}(E)$ is a monotonically decreasing function. From Eq.~\eqref{eq:IFVapp}, it also follows that $I_{\rm FV}(E)$ has a simple pole at $E=E_{\bm{n}}$. Near the pole energy, a single term $\propto 1/(E-E_{\bm{n}})$ dominates, so we obtain for an infinitesimal $\epsilon>0$,
\begin{align}
   I_{\rm FV}(E_{\bm{n}}\pm \epsilon)
   &= \pm\frac{1}{L^{3}\epsilon} .
   \label{eq:infini}
\end{align}
Combining Eqs.~\eqref{eq:IFVprime} and \eqref{eq:infini}, we find that  there must be a zero of $I_{\rm FV}(E)$ in between two neighboring poles.

\begin{figure}[tbp]
    \centering
    \includegraphics[width=7cm]{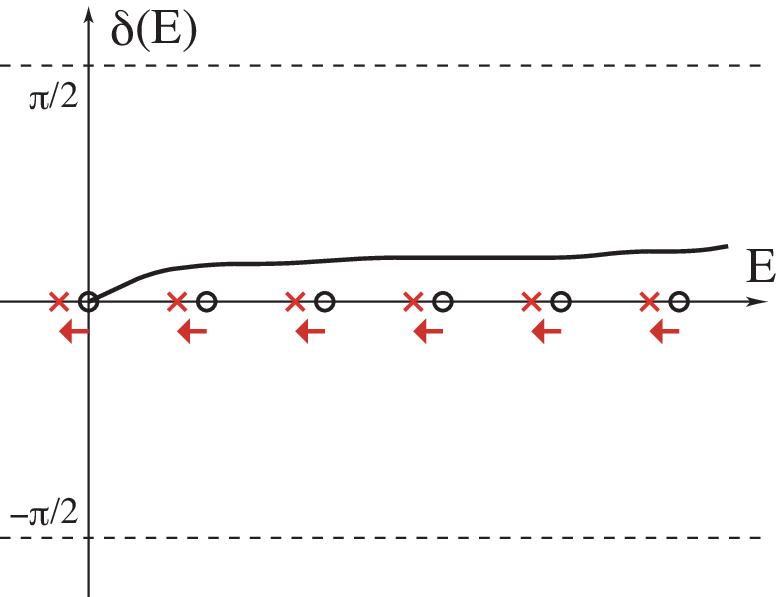}
    \includegraphics[width=7cm]{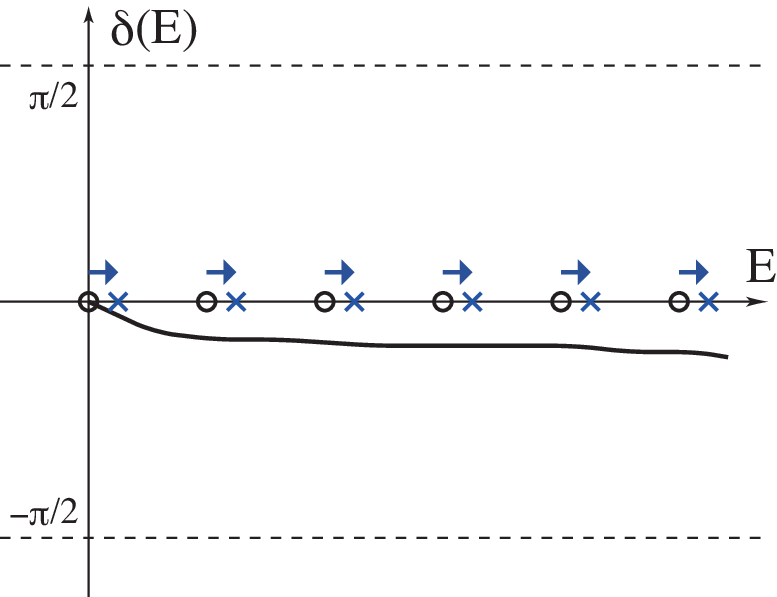}
    \caption{\label{fig:phase}
    Schematic illustration of the infinite-volume phase shift (solid lines), the eigenenergies of the free Hamiltonian in finite volume (circles), and the eigenenergies of the full Hamiltonian in finite volume (crosses).}
\end{figure}

% shifted scattering states
Now we study the eigenenergy for a given phase shift. First, consider the noninteracting case with $\delta=0$ in the relevant energy region. In this case, $\cot\delta\to \infty$, so Eq.~\eqref{eq:Luescher} indicates that $I(E_{\rm FV})\to \infty$. Because of the poles of $I_{\rm FV}(E)$, this gives the noninteracting eigenenergies
\begin{align}
   E_{\rm FV}
   &= E_{\bm{n}}
   \quad (\text{noninteracting}) ,
\end{align}
as expected. Next, we consider a weakly interacting case with $0< |\delta|\ll 1$ for some energy region. The eigenenergies in this region are slightly shifted as $E_{\rm FV}= E_{\bm{n}}+\Delta E_{\bm{n}}$. If $\Delta E_{\bm{n}}$ is sufficiently smaller than the level spacing from the nearest $E_{\bm{n}}$, we use Eq.~\eqref{eq:infini} to obtain
\begin{align}
   \frac{\sqrt{2\mu E_{\rm FV}}}{\delta}
   &= -\frac{2}{\mu L^{3}\Delta E_{\bm{n}}} ,
\end{align}
which means that 
\begin{align}
   E_{\rm FV}
   &= E_{\bm{n}}-\frac{\sqrt{2}}{\mu L^{2}|\bm{n}|}\delta
   \quad (\text{weakly interacting}) .
   \label{eq:weakint}
\end{align}
In other words, for a weakly attractive (repulsive) phase shift $\delta>0$ ($\delta<0$), the eigenenergy decreases (increases) from the noninteracting value (see Fig.~\ref{fig:phase}). In fact, this behavior generally holds for a finite $\delta$ and $\Delta E_{\bm{n}}$, when the sign of $\delta$ is unchanged. This is because the magnitude of the energy shift $|\Delta E_{\bm{n}}|$ is bounded by the zero of $I_{\rm FV}(E)$. Thus, as long as the sign of the phase shift is kept fixed in a given energy region, the number of eigenstates coincides with that of the noninteracting scattering states and there is no energy level in addition to the shifted scattering states.

%--figure---------------------------------
%--figure---------------------------------

\begin{figure}[tbp]
    \centering
    \includegraphics[width=7cm]{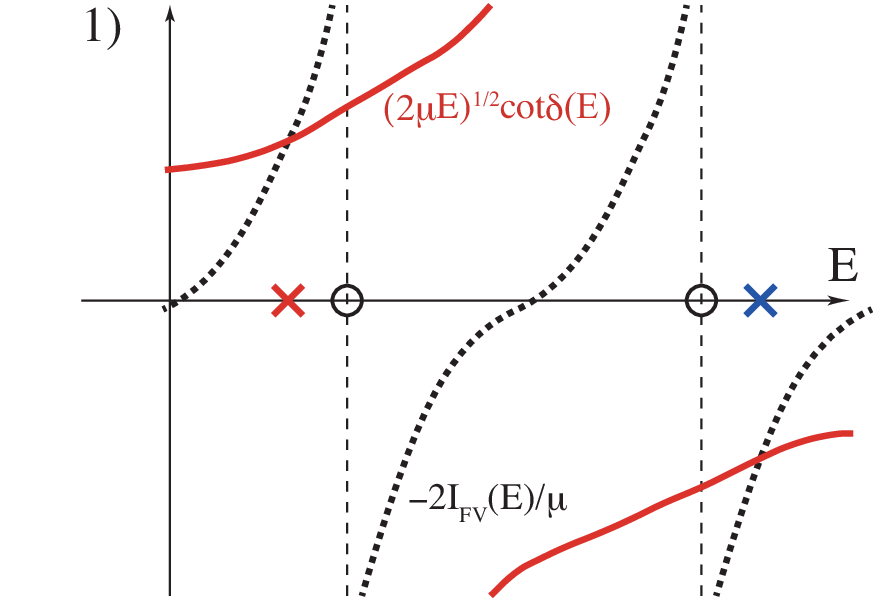}
    \includegraphics[width=7cm]{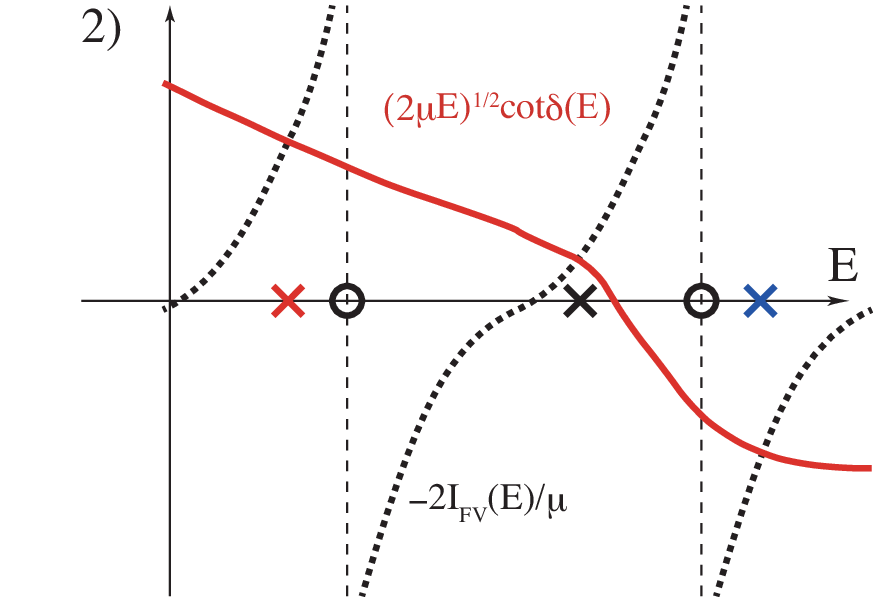}
    \caption{\label{fig:phaseres}
    Schematic illustration of the behaviors of $\sqrt{2\mu E}\cot\delta(E)$ (solid lines) and $-2I_{\rm FV}(E)/\mu$ (dotted lines) as functions of $E$ with two cases explained in the text.
    The eigenenergies of the free (full) Hamiltonian is denoted by the circles (crosses).}
\end{figure}

% sign change
Now we consider that $\delta$ changes the sign from positive to negative between two neighboring eigenenergies of the free Hamiltonian, $E_{\bm{n}}^{1}$ and $E_{\bm{n}}^{2}>E_{\bm{n}}^{1}$ as
\begin{align}
   \delta(E_{\bm{n}}^{1})
   &>0, \quad
   \delta(E_{\bm{n}}^{2})<0 .
\end{align}
To make the situation clear, we assume that $L$ is sufficiently large and the phase shift varies monotonically in the region $E_{\bm{n}}^{1}<E<E_{\bm{n}}^{2}$. According to Eq.~\eqref{eq:weakint}, the full eigenenergy is shifted down (up) from $E_{\bm{n}}^{1}$ ($E_{\bm{n}}^{2}$). The behavior of $\delta(E)$ between  $E_{\bm{n}}^{1}$ and $E_{\bm{n}}^{2}$ can be classified into two cases:
\begin{itemize}
\item [(1)] $d\delta(E)/dE<0$ and the phase shift changes the sign at $\delta=0$.
\item [(2)] $d\delta(E)/dE>0$ and the phase shift changes the sign at $\delta=\pi/2$.
\end{itemize}
Case (1) corresponds to the smooth change of the sign of the interaction from attractive to repulsive, while case (2) corresponds to the existence of a sharp resonance. Noting that $\cot\delta\to \infty$ at $\delta\to 0$ and $\cot\delta= 0$ at $\delta=\pi/2$, we expect the behaviors of $\sqrt{2\mu E}\cot\delta(E)$ in cases (1) and (2) as shown in Fig.~\ref{fig:phaseres}. On the other hand, from Eq.~\eqref{eq:IFVprime}, $-2I_{\rm FV}(E)/\mu$ is a monotonically increasing function. Because the finite-volume eigenenergy is obtained by Eq.~\eqref{eq:Luescher}, we find that there is no eigenstate in the region $E_{\bm{n}}^{1}<E<E_{\bm{n}}^{2}$ for case (1), except for very rapid energy dependence of $\delta(E)$. For case (2), the function
\begin{align}
   f(E)
   \equiv \sqrt{2\mu E}\cot\delta(E)
   -\left[-\frac{2}{\mu }I_{\rm FV}(E)\right]
\end{align}
is continuous and monotonic in $E_{\bm{n}}^{1}<E<E_{\bm{n}}^{2}$  and $f(E_{\bm{n}}^{1}+\epsilon)<0<f(E_{\bm{n}}^{2}-\epsilon)$. Thus, there must be $f(E_{r})=0$ in $E_{\bm{n}}^{1}<E_{r}<E_{\bm{n}}^{2}$ by the intermediate value theorem. This means that, if and only if the phase shift crosses $\delta=\pi/2$ in infinite volume, there appears one finite-volume eigenstate in addition to those shifted from the noninteracting eigenstates. When the phase shift crosses $\pi/2$ twice in the relevant energy region, two additional energy levels appear in finite volume. In this way, the number of additional eigenstates in finite volume is determined by the behavior of the phase shift on the real energy axis, rather than the number of poles in the complex energy plane.

%--figure---------------------------------
%--figure---------------------------------

% Lambda(1405)
In the case of $\Lambda(1405)$, the phase shift of the $\pi\Sigma$ scattering crosses $\pi/2$ only once between the $\bar{K}N$ and $\pi\Sigma$ thresholds, although there are two complex poles in the $\bar{K}N$-$\pi\Sigma$ energy region (see Sec.~\ref{subsec:L1405}). Thus, in finite volume, the number of energy levels representing the resonance in addition to the scattering states is 1. We note that it is possible to increase the accuracy of the determination of the scattering amplitude by the use of the asymmetric box and the moving frame~\cite{MartinezTorres:2012yi}. The twisted boundary conditions also allows one to determine the phase shift at different eigenmomenta~\cite{Ozaki:2012ce}. Even in these cases, what is determined by the finite-volume eigenenergies is the phase shift on the real energy axis, and the amplitude in the complex energy plane is accessible only through the analytic continuation, which requires a parametrization of the amplitude. To pin down the pole structure in the complex energy plane, a detailed analysis of the system is required, as discussed in Ref.~\cite{MartinezTorres:2012yi}.


%\bibliographystyle{apsrev4-1}
%\bibliographystyle{h-physrev4}
%\bibliography{refs}

\begin{thebibliography}{10}

\bibitem{Olive:2016xmw}
C.~Patrignani {\em et~al.} (Particle Data Group),
\newblock Chin. Phys. {\bf C40}, 100001 (2016).
%%CITATION = CHPHD,C40,100001;%%

\bibitem{Brambilla:2010cs}
N.~Brambilla {\em et~al.},
\newblock Eur. Phys. J. C {\bf 71}, 1534 (2011).
%%CITATION = ARXIV:1010.5827;%%

\bibitem{Hosaka:2016pey}
A.~Hosaka, T.~Iijima, K.~Miyabayashi, Y.~Sakai and S.~Yasui,
\newblock PTEP {\bf 2016}, 062C01 (2016).
%%CITATION = ARXIV:1603.09229;%%

\bibitem{Dalitz:1959dn}
R.~H. Dalitz and S.~F. Tuan,
\newblock Phys. Rev. Lett. {\bf 2}, 425 (1959).
%%CITATION = PRLTA,2,425;%%

\bibitem{Dalitz:1960du}
R.~H. Dalitz and S.~F. Tuan,
\newblock Annals Phys. {\bf 10}, 307 (1960).
%%CITATION = APNYA,10,307;%%

\bibitem{Kaiser:1995eg}
N.~Kaiser, P.~B. Siegel and W.~Weise,
\newblock Nucl. Phys. A {\bf 594}, 325 (1995).
%%CITATION = NUCL-TH 9505043;%%

\bibitem{Oset:1998it}
E.~Oset and A.~Ramos,
\newblock Nucl. Phys. A {\bf 635}, 99 (1998).
%%CITATION = NUCL-TH 9711022;%%

\bibitem{Oller:2000fj}
J.~A. Oller and U.~G. Meissner,
\newblock Phys. Lett. B {\bf 500}, 263 (2001).
%%CITATION = HEP-PH 0011146;%%

\bibitem{Jido:2003cb}
D.~Jido, J.~A. Oller, E.~Oset, A.~Ramos and U.~G. Meissner,
\newblock Nucl. Phys. A {\bf 725}, 181 (2003).
%%CITATION = NUCL-TH 0303062;%%

\bibitem{Hyodo:2007jq}
T.~Hyodo and W.~Weise,
\newblock Phys. Rev. C {\bf 77}, 035204 (2008).
%%CITATION = ARXIV:0712.1613;%%

\bibitem{Hyodo:2011ur}
T.~Hyodo and D.~Jido,
\newblock Prog. Part. Nucl. Phys. {\bf 67}, 55 (2012).
%%CITATION = ARXIV:1104.4474;%%

\bibitem{Kamiya:2016jqc}
Y.~Kamiya, K. Miyahara, S. Ohnishi, Y. Ikeda, T. Hyodo, E. Oset, and W. Weise,
\newblock Nucl. Phys. A {\bf 954}, 41 (2016).
%%CITATION = ARXIV:1602.08852;%%

\bibitem{Weinstein:1982gc}
J.~D. Weinstein and N.~Isgur,
\newblock Phys. Rev. Lett. {\bf 48}, 659 (1982).
%%CITATION = PRLTA,48,659;%%

\bibitem{Weinstein:1983gd}
J.~D. Weinstein and N.~Isgur,
\newblock Phys. Rev. D {\bf 27}, 588 (1983).
%%CITATION = PHRVA,D27,588;%%

\bibitem{Oller:1997ti}
J.~A. Oller and E.~Oset,
\newblock Nucl. Phys. A {\bf 620}, 438 (1997).
%%CITATION = HEP-PH 9702314;%%

\bibitem{Oller:1997ng}
J.~A. Oller, E.~Oset and J.~R. Pelaez,
\newblock Phys. Rev. Lett. {\bf 80}, 3452 (1998).
%%CITATION = HEP-PH 9803242;%%

\bibitem{Oller:1998hw}
J.~A. Oller, E.~Oset and J.~R. Pelaez,
\newblock Phys. Rev. D {\bf 59}, 074001 (1999).
%%CITATION = HEP-PH 9804209;%%

\bibitem{GomezNicola:2001as}
A.~Gomez~Nicola and J.~R. Pelaez,
\newblock Phys. Rev. D {\bf 65}, 054009 (2002).
%%CITATION = HEP-PH/0109056;%%

\bibitem{Baru:2003qq}
V.~Baru, J.~Haidenbauer, C.~Hanhart, Y.~Kalashnikova and A.~E. Kudryavtsev,
\newblock Phys. Lett. B {\bf 586}, 53 (2004).
%%CITATION = HEP-PH/0308129;%%

\bibitem{Pelaez:2015qba}
J.~R. Pelaez,
\newblock Phys. Rept. {\bf 658}, 1 (2016).
%%CITATION = ARXIV:1510.00653;%%

\bibitem{Weinberg:1965zz}
S.~Weinberg,
\newblock Phys. Rev. {\bf 137}, B672 (1965).
%%CITATION = PHRVA,137,B672;%%

\bibitem{Baru:2010ww}
V.~Baru, C.~Hanhart, {\relax Yu}.~S. Kalashnikova, A.~E. Kudryavtsev and A.~V.
  Nefediev,
\newblock Eur. Phys. J. A {\bf 44}, 93 (2010).
%%CITATION = ARXIV:1001.0369;%%

\bibitem{Hyodo:2011qc}
T.~Hyodo, D.~Jido and A.~Hosaka,
\newblock Phys. Rev. C {\bf 85}, 015201 (2012).
%%CITATION = ARXIV:1108.5524;%%

\bibitem{Aceti:2012dd}
F.~Aceti and E.~Oset,
\newblock Phys. Rev. D {\bf 86}, 014012 (2012).
%%CITATION = ARXIV:1202.4607;%%

\bibitem{Hyodo:2013nka}
T.~Hyodo,
\newblock Int. J. Mod. Phys. A {\bf 28}, 1330045 (2013).
%%CITATION = ARXIV:1310.1176;%%

\bibitem{Hyodo:2013iga}
T.~Hyodo,
\newblock Phys. Rev. Lett. {\bf 111}, 132002 (2013).
%%CITATION = ARXIV:1305.1999;%%

\bibitem{Hyodo:2014bda}
T.~Hyodo,
\newblock Phys. Rev. C {\bf 90}, 055208 (2014).
%%CITATION = ARXIV:1407.2372;%%

\bibitem{Sekihara:2014kya}
T.~Sekihara, T.~Hyodo and D.~Jido,
\newblock PTEP {\bf 2015}, 063D04 (2015).
%%CITATION = ARXIV:1411.2308;%%

\bibitem{Aceti:2014ala}
F.~Aceti, L.~Dai, L.~Geng, E.~Oset and Y.~Zhang,
\newblock Eur. Phys. J. A {\bf 50}, 57 (2014).
%%CITATION = ARXIV:1301.2554;%%

\bibitem{Garcia-Recio:2015jsa}
C.~Garcia-Recio, C.~Hidalgo-Duque, J.~Nieves, L.~L. Salcedo and L.~Tolos,
\newblock Phys. Rev. D {\bf 92}, 034011 (2015).
%%CITATION = ARXIV:1506.04235;%%

\bibitem{Guo:2015daa}
Z.-H. Guo and J.~A. Oller,
\newblock Phys. Rev. D {\bf 93}, 096001 (2016).
%%CITATION = ARXIV:1508.06400;%%

\bibitem{Kamiya:2015aea}
Y.~Kamiya and T.~Hyodo,
\newblock Phys. Rev. C {\bf 93}, 035203 (2016).
%%CITATION = ARXIV:1509.00146;%%

\bibitem{Sekihara:2015gvw}
T.~Sekihara, T.~Arai, J.~Yamagata-Sekihara and S.~Yasui,
\newblock Phys. Rev. C {\bf 93}, 035204 (2016).
%%CITATION = ARXIV:1511.01200;%%

\bibitem{Guo:2016wpy}
Z.-H. Guo and J.~A. Oller,
\newblock Phys. Rev. D {\bf 93}, 054014 (2016).
%%CITATION = ARXIV:1601.00862;%%

\bibitem{Kamiya:2016oao}
Y.~Kamiya and T.~Hyodo,
\newblock PTEP {\bf 2017}, 023 (2017).
%%CITATION = ARXIV:1607.01899;%%

\bibitem{Kukulin}
V.~I. Kukulin, V.~M. Krasnopol'sky and J.~Horacek,
\newblock {\em Theory of Resonances} (Kluwer Academic Publishers, Dordrecht,
  1989).

\bibitem{Moiseyev}
N.~Moiseyev,
\newblock {\em Non-Hermitian Quantum Mechanics} (Cambridge University Press,
  Cambridge, 2011).

\bibitem{Luscher:1985dn}
M.~Luscher,
\newblock Commun. Math. Phys. {\bf 104}, 177 (1986).
%%CITATION = CMPHA,104,177;%%

\bibitem{Luscher:1986pf}
M.~Luscher,
\newblock Commun. Math. Phys. {\bf 105}, 153 (1986).
%%CITATION = CMPHA,105,153;%%

\bibitem{Luscher:1990ux}
M.~Luscher,
\newblock Nucl. Phys. B {\bf 354}, 531 (1991).
%%CITATION = NUPHA,B354,531;%%

\bibitem{Beane:2003da}
S.~Beane, P.~Bedaque, A.~Parreno and M.~Savage,
\newblock Phys. Lett. B {\bf 585}, 106 (2004).
%%CITATION = HEP-LAT/0312004;%%

\bibitem{Koma:2004wz}
Y.~Koma and M.~Koma,
\newblock Nucl. Phys. B {\bf 713}, 575 (2005).
%%CITATION = HEP-LAT 0406034;%%

\bibitem{Lage:2009zv}
M.~Lage, U.-G. Meissner and A.~Rusetsky,
\newblock Phys. Lett. B {\bf 681}, 439 (2009).
%%CITATION = 0905.0069;%%

\bibitem{Doring:2011ip}
M.~Doring, J.~Haidenbauer, U.-G. Meissner and A.~Rusetsky,
\newblock Eur. Phys. J. A {\bf 47}, 163 (2011).
%%CITATION = ARXIV:1108.0676;%%

\bibitem{Doring:2011vk}
M.~Doring, U.-G. Meissner, E.~Oset and A.~Rusetsky,
\newblock Eur. Phys. J. A {\bf 47}, 139 (2011).
%%CITATION = 1107.3988;%%

\bibitem{MartinezTorres:2011pr}
A.~Martinez~Torres, L.~R.~Dai, C.~Koren, D.~Jido and E.~Oset,
\newblock Phys. Rev. D {\bf 85}, 014027 (2012).
%%CITATION = ARXIV:1109.0396;%%

\bibitem{MartinezTorres:2012yi}
A.~Martinez~Torres, M.~Bayar, D.~Jido and E.~Oset,
\newblock Phys. Rev. C {\bf 86}, 055201 (2012).
%%CITATION = ARXIV:1202.4297;%%

\bibitem{Hall:2013qba}
J.~M.~M. Hall, A.~C.~-P. Hsu, D.~B. Leinweber, A.~W. Thomas and R.~D. Young,
\newblock Phys. Rev. {\bf D87}, 094510 (2013).
%%CITATION = ARXIV:1303.4157;%%

\bibitem{Albaladejo:2016jsg}
M.~Albaladejo, P.~Fernandez-Soler and J.~Nieves,
\newblock Eur. Phys. J. C {\bf 76}, 573 (2016).
%%CITATION = ARXIV:1606.03008;%%

\bibitem{Menadue:2011pd}
B.~J. Menadue, W.~Kamleh, D.~B. Leinweber and M.~S. Mahbub,
\newblock Phys. Rev. Lett. {\bf 108}, 112001 (2012).
%%CITATION = ARXIV:1109.6716;%%

\bibitem{Hall:2014uca}
J.~M.~M. Hall, W.~Kamleh, D.~B.~Leinweber, B.~J.~Menadue, B.~J.~Owen, A.~W.~Thomas, R.~D.~Young,
\newblock Phys. Rev. Lett. {\bf 114}, 132002 (2015).
%%CITATION = ARXIV:1411.3402;%%

\bibitem{Molina:2015uqp}
R.~Molina and M.~Doring,
\newblock Phys. Rev. D {\bf 94}, 056010 (2016),
\newblock [Addendum: Phys. Rev. D {\bf 94}, 079901 (2016)].
%%CITATION = ARXIV:1512.05831;%%

\bibitem{Liu:2016wxq}
Z.-W. Liu, J.~M.~M. Hall, D.~B. Leinweber, A.~W. Thomas and J.-J. Wu,
\newblock Phys. Rev. D {\bf 95}, 014506 (2017).
%%CITATION = ARXIV:1607.05856;%%

\bibitem{Sekihara:2012xp}
T.~Sekihara and T.~Hyodo,
\newblock Phys. Rev. C {\bf 87}, 045202 (2013).
%%CITATION = ARXIV:1209.0577;%%

\bibitem{Feshbach:1958nx}
H.~Feshbach,
\newblock Ann. Phys. {\bf 5}, 357 (1958).
%%CITATION = APNYA,5,357;%%

\bibitem{Feshbach:1962ut}
H.~Feshbach,
\newblock Ann. Phys. {\bf 19}, 287 (1962).
%%CITATION = APNYA,19,287;%%

\bibitem{Chew:1960iv}
G.~F. Chew and S.~Mandelstam,
\newblock Phys. Rev. {\bf 119}, 467 (1960).
%%CITATION = PHRVA,119,467;%%

\bibitem{Bjorken:1960zz}
J.~D. Bjorken,
\newblock Phys. Rev. Lett. {\bf 4}, 473 (1960).
%%CITATION = PRLTA,4,473;%%

\bibitem{Ikeda:2011pi}
Y.~Ikeda, T.~Hyodo and W.~Weise,
\newblock Phys. Lett. B {\bf 706}, 63 (2011).
%%CITATION = ARXIV:1109.3005;%%

\bibitem{Ikeda:2012au}
Y.~Ikeda, T.~Hyodo and W.~Weise,
\newblock Nucl. Phys. A {\bf 881}, 98 (2012).
%%CITATION = ARXIV:1201.6549;%%

\bibitem{Sekihara:2016xnq}
T.~Sekihara,
\newblock Phys. Rev. C {\bf 95}, 025206 (2017).
%%CITATION = ARXIV:1609.09496;%%

\bibitem{Miyahara}
K.~Miyahara, T.~Hyodo and W.~Weise,
\newblock arXiv:1804.08269 [nucl-th].

\bibitem{Bazzi:2011zj}
SIDDHARTA Collaboration, M.~Bazzi {\em et~al.},
\newblock Phys. Lett. B {\bf 704}, 113 (2011).
%%CITATION = ARXIV:1105.3090;%%

\bibitem{Bazzi:2012eq}
SIDDHARTA Collaboration, M.~Bazzi {\em et~al.},
\newblock Nucl. Phys. A {\bf 881}, 88 (2012).
%%CITATION = ARXIV:1201.4635;%%

\bibitem{Meissner:2004jr}
U.~G. Meissner, U.~Raha and A.~Rusetsky,
\newblock Eur. Phys. J. C {\bf 35}, 349 (2004).
%%CITATION = HEP-PH/0402261;%%

\bibitem{Castillejo:1956ed}
L.~Castillejo, R.~H. Dalitz and F.~J. Dyson,
\newblock Phys. Rev. {\bf 101}, 453 (1956).
%%CITATION = PHRVA,101,453;%%

\bibitem{Yamaguchi:2016kxa}
Y.~Yamaguchi and T.~Hyodo,
\newblock Phys. Rev. C {\bf 94}, 065207 (2016).
%%CITATION = ARXIV:1607.04053;%%

\bibitem{delaMadrid:2002cz}
R.~de~la Madrid and M.~Gadella,
\newblock Am. J. Phys. {\bf 70}, 626 (2002).
%%CITATION = QUANT-PH/0201091;%%

\bibitem{Ozaki:2012ce}
S.~Ozaki and S.~Sasaki,
\newblock Phys. Rev. D {\bf 87}, 014506 (2013).
%%CITATION = ARXIV:1211.5512;%%

\end{thebibliography}

\end{document}